\documentclass[twocolumn,showpacs,prx,superscriptaddress,amsmath,amssymb]{revtex4}

\usepackage{booktabs}
\usepackage{color}
\usepackage{graphicx}% Include figure files
\usepackage{dcolumn}% Align table columns on decimal point
\usepackage{bm}% bold math
\usepackage{graphicx}
\begin{document}

\title{Relativistic Impulse Approximation in Compton Scattering}

\author{Chen-Kai Qiao}
\affiliation{College of Physics, Sichuan University, Chengdu, Sichuan, 610065}

\author{Hsin-Chang Chi}
% \email{hsinchang@mail.ndhu.edu.tw}
\affiliation{Department of Physics, National Dong Hwa University, Shoufeng, Hualien, 97401}

\author{Lei Zhang}
\affiliation{College of Physics, Sichuan University, Chengdu, Sichuan, 610065}

\author{Peng Gu}
\affiliation{College of Physics, Sichuan University, Chengdu, Sichuan, 610065}

\author{\\Cheng-Pang Liu}
\affiliation{Department of Physics, National Dong Hwa University, Shoufeng, Hualien, 97401}

\author{Chang-Jian Tang}
\affiliation{College of Physics, Sichuan University, Chengdu, Sichuan, 610065}

\author{Shin-Ted Lin}
\email{stlin@scu.edu.cn}
\affiliation{College of Physics, Sichuan University, Chengdu, Sichuan, 610065}
\affiliation{Institute of Physics, Academia Sinica, Taipei, 11529}

\author{Keh-Ning Huang}
\email{knhuang1206@gmail.com}
\affiliation{Institute of Atomic and Molecular Physics, Sichuan University,
Chengdu, Sichuan, 610065}
\affiliation{Department of Physics, National Taiwan University, Taipei, 10617}

\date{\today}

\begin{abstract}
Relativistic impulse approximation (RIA) has been widely used in atomic, condensed matter, nuclear, and elementary particle physics. In former treatments of RIA formulation, differential cross sections for Compton scattering processes were factorized into atomic Compton profiles by performing further simplified approximations in the integration. In this study, we develop an ``exact'' numerical method without using any further simplified approximations or factorization treatments. The validity of the approximations and factorizations used in former RIA treatments can be tested using our approach. Calculations for C, Cu, Ge, and Xe atomic systems are carried out using Dirac-Fock wavefunctions, and comparisons between the proposed approach and former treatments of RIA are performed and discussed in detail. Numerical results indicate that these simplified approximations work reasonably in the Compton peak region, and our results have little difference with the best of the former RIA treatments in the entire energy region. While in regions far from the Compton peak, the RIA results become inaccurate, even when our ``exact'' numerical treatment is used.

\vspace{2pc}
\end{abstract}


\maketitle

%\tableofcontents

\section{Introduction\label{sec:1}}

Atomic Compton scatterings, which have been widely investigated over the past few decades, are expressed as follows:
\begin{equation}
\hbar\omega_{i}+ A \longrightarrow \hbar\omega_{f}+e^{-}+ A^{+}
\end{equation}
Many aspects of physics, such as electron correlations \cite{Kubo,Pisani}, electron momentum distributions \cite{Cooper0,Cooper,Aguiar}, Fermi surfaces \cite{Wang}, X-ray, and gamma-ray radiations \cite{Pratt,Porter,Phuoc}, have been revealed through them. Moreover, Compton scatterings have been utilized to develop the modern gamma-ray spectrometer and imaging devices \cite{Takada,Mihailescu,Chiu}.

For convenience, Compton scattering is conventionally approached using the Klein-Nishina formula from free electron approximation (FEA) \cite{Klein-Nishina,Sakurai}. In FEA, electron interactions with atomic ions are neglected, and electrons are also assumed to be at rest prior to photon scatterings in the laboratory frame. In the Klein-Nishina formula, the energy of the scattered photon $\omega_{C}$ is completely determined using its scattering angle $\theta$ as follows:
\begin{equation}
\omega_{C}=\frac{\omega_{i}}{1+\omega_{i}(1-\cos\theta)/mc^{2}}
\end{equation}
The Klein-Nishina formula works perfectly in high-energy regions, where the electrons are asymptotical free. However, in low-energy regions, where the atomic binding effects are present, the FEA becomes inappropriate and the Klein-Nishina formula fails to explain the experiments \cite{Pratt}.

\begin{figure}
\includegraphics[width=0.525\textwidth]{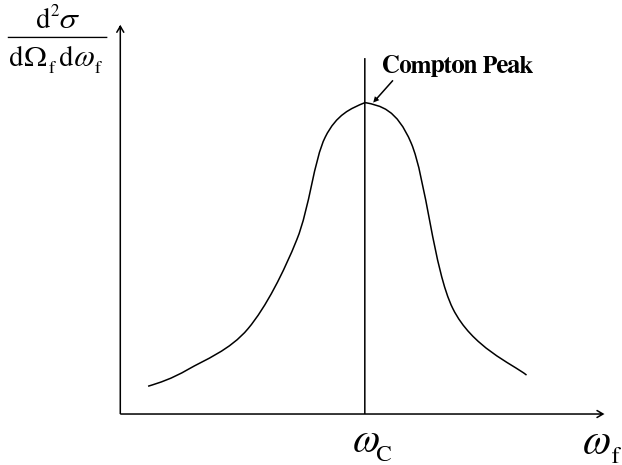}
\caption{Compton spectrum in the impulse approximation (IA) model at the scattering angle $\theta$.} %ͼÌâ
\label{Energy Spectrum}
\end{figure}

The atomic binding effects are systematically treated in impulse approximation (IA) \cite{Eisenberger1,Eisenberger2,Ribberfors1,Ribberfors2,Ribberfors3,Ribberfors4}, in which the electrons in an atom have a momentum distribution. The motion of electrons causes a Doppler broadened Compton spectrum, as shown in Fig \ref{Energy Spectrum}. In the former treatments of IA models, the doubly-differential cross section (DDCS) of Compton scatterings can be factorized into two parts, given by
\begin{equation}
\bigg(\frac{d^{2}\sigma}{d\omega_{f}d\Omega_{f}}\bigg)_{IA} = Y \cdot J \label{IA}
\end{equation}
Here, $Y$ is a factor dependent on kinematical and dynamical properties of Compton scatterings, and irrelevant to the electronic structure of target materials. The correction factor $J$, known as the Compton profile, is related to the momentum distributions of electrons in the atomic or molecular ground state.

Currently, former IA treatments, which incorporate the factorization in Eq. (\ref{IA}), are widely applied in interdisciplinary studies, particularly in condensed matter, nuclear, and elementary particle physics. Sophisticated electronic structures \cite{Gillet,Sahariya}, electron correlations \cite{Kubo,Pisani}, band structures, and Fermi surfaces \cite{Wang,Rathor} in condensed matter physics are studied using Compton profiles. The current Geant4 and other Monte Carlo simulation packages in nuclear and particle physics adopt the IA formulation and the Compton profiles \cite{Brusa,Salvat,Brown}. The conclusions of these interdisciplinary studies depend strictly on the validity of factorization in Eq. (\ref{IA}). Previously, it was believed that this factorization result adopted in former relativistic impulse approximation (RIA) treatments does not essentially change the physical results \cite{Ribberfors1}. However, this assumption has not been quantitatively analysed in the past years. Thus, this study is focused on clarifying whether essential differences in IA formulations exist with and without these factorization treatments.

Therefore, for a comprehensive study of atomic Compton Scattering processes, in this study, we develop an ``exact'' numerical treatment of relativistic impulse approximation (RIA) without invoking the factorization in Eq. (\ref{IA}). Then we apply the present approach to Compton scattering with several atomic systems, and the results are compared with those of former treatments of RIA. Furthermore, a careful analysis of the adequacy of former RIA treatments and the validity of factorization in Eq. (\ref{IA}) is provided in this work. Moreover, effective Compton profiles are proposed and analysed to quantify the differences between our results and those of former RIA treatments.

Recently, LaJohn compared various treatments of RIA formulation in a similar manner, and achieved the nonrelativistic limit of RIA for low-momentum-transfer cases \cite{LaJohn}. However, his work is limited to hydrogen-like systems. In our study, more complicated atomic systems are considered. We apply the present scheme to the atoms C, Cu, Ge, and Xe, which are chosen to represent elements in the small-$Z$, middle-$Z$, and large-$Z$ regimes. To obtain the ground state wavefunctions for atomic systems, we have employed the fully relativistic Dirac-Fock theory \cite{Grant1961,Desclaux1971,Desclaux,Grant,Ankudinov,Visscher}. In the Dirac-Fock formalism, electrons in atomic systems are quantized and many-body effects, including electron exchange and electron correlation interactions, are effectively considered.

Recently, there has been great interest in experimentally detecting dark matter particles \cite{Undagoitia,CDEX0,CDEX,CDMS,PandaX,LUX,XENON} and neutrino-less double beta decays \cite{Rodejohann,GERDA,GERDA2,EXO,KamLAND-Zen}. These experiments, which utilise high-purity Germanium and Xenon detectors, require a sufficiently low radiation background. Compton scattering is one of the most dominant radiation backgrounds for X-ray and gamma rays which must be suppressed and subtracted. Therefore, studying the atomic Compton scattering effects in detectors could have a great impact on these elementary particle experiments. Recent studies using former treatments of RIA have indicated that low momentum transfer Compton scattering plays a remarkable role in dark matter direct detections \cite{Barker,Ramanathan}. Further, our method can be easily applied to this area, and could impact and guide the analysis and subtraction of Compton scattering backgrounds in particle physics experiments.

This paper is organized as follows: Section \ref{sec:2} introduces the RIA formulation, and is divided into two subsections. In Section \ref{sec:2a}, we briefly review the former treatments of Compton scattering in RIA formulation. In Section \ref{sec:2b}, we describe our present numerical treatment of RIA for application to atomic Compton scatterings. The results and comparisons of our approach and former RIA treatments are presented in Section \ref{sec:3}. Finally, the conclusions and future perspectives are provided in Section \ref{sec:4}.

\section{Relativistic Impulse Approximation \label{sec:2}}

\subsection{Former Treatments \label{sec:2a}}

In this section, we give a theoretical description of the former treatments of RIA formulations for Compton scatterings. The nonrelativistic impulse approximation approach can be derived similar to the relativistic case.

In the RIA formulations, consider an incident photon with energy $\omega_{i}$ and momentum $\boldsymbol{k}_{i}$ scattering with an electron which has energy $E_{i}$ and momentum $\boldsymbol{p}_{i}$. After scattering, the energy and momentum of emitted photon are $\omega_{f}$ and $\boldsymbol{k}_{f}$, and energy and momentum of final state electron are $E_{f}$ and $\boldsymbol{p}_{f}$. Then the DDCS of Compton scattering in RIA formulation is given by \cite{Ribberfors1,Ribberfors2,Ribberfors3}
\begin{eqnarray}
\frac{d^{2}\sigma}{d\omega_{f}d\Omega_{f}} & = & \frac{r_{0}^{2}m^{2}c^{4}}{2} \frac{\omega_{f}}{\omega_{i}} \iiint{d^{3}p_{i}\rho(\boldsymbol{p}_{i})
                                                 \frac{X(K_{i},K_{f})}{E_{i}E_{f}}} \nonumber
\\
                                           &   & \times \ \delta (E_{i}+\omega_{i}-E_{f}-\omega_{f})
\label{doubly differential cross section1}
\end{eqnarray}
where $r_{0}$ is the electron classical charge radius, functions $K_{i}$, $K_{f}$ are defined as
\begin{eqnarray}
K_{i} & = & k_{i}^{\mu}\cdot p_{i\mu}=\frac{E_{i}\cdot \omega_{i}}{c^{2}}-\boldsymbol{p}_{i} \cdot \boldsymbol{k}_{i}
\\
K_{f} & = & k_{f}^{\mu}\cdot p_{i\mu}=\frac{E_{i}\cdot \omega_{f}}{c^{2}}-\boldsymbol{p}_{i} \cdot \boldsymbol{k}_{f} \nonumber
\\
      & = & K_{i}-\frac{\omega_{i}\omega_{f}(1-\cos\theta)}{c^{2}}
\end{eqnarray}
and the kernel function $X(K_{i},K_{f})$ is defined as
\begin{eqnarray}
X(K_{i},K_{f}) & = & \frac{K_{i}}{K_{f}}+\frac{K_{f}}{K_{i}}
                     +2m^{2}c^{2}
                     \bigg(
                       \frac{1}{K_{i}}-\frac{1}{K_{f}}
                     \bigg) \nonumber
\\
               &   & +m^{4}c^{4}
                     \bigg(
                       \frac{1}{K_{i}}-\frac{1}{K_{f}}
                     \bigg)^{2} \label{function X}
\end{eqnarray}
Here, $\rho(\boldsymbol{p}_{i})$ denotes the momentum distribution of electrons, which is calculated through
\begin{equation}
\rho(\boldsymbol{p}_{i})=\sum_{a}|\phi_{a}(\boldsymbol{p}_{i})|^{2}
\end{equation}
Here, the sum is over all electrons, and $\phi_{a}(\boldsymbol{p}_{i})$ is the momentum wavefunction for \emph{a}-th electron, which is related to the electron's position wavefunction $\psi_{a}(\boldsymbol{r})$ through Fourier transformation
\begin{equation}
\phi_{a}(\boldsymbol{p}_{i})=\frac{1}{(2\pi)^{3/2}} \int{d^{3}r\psi_{a}(\boldsymbol{r})e^{i\boldsymbol{p}_{i}\cdot\boldsymbol{r}}}
\end{equation}
In this work, we employ the fully relativistic Dirac-Fock theory to calculate the total wavefunctions for atomic ground states and the wavefunctions for individual electrons.

\begin{figure}
\includegraphics[width=0.525\textwidth]{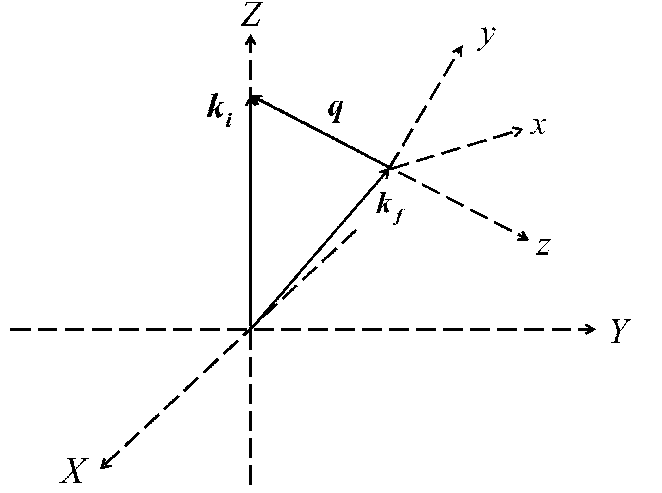}
\caption{Coordinate system $XYZ$ and $xyz$. Coordinate system $XYZ$ is chosen such that the $Z$ axis is along the direction of initial photon $\boldsymbol{k}_{i}$, and $X$ axis can be chosen as arbitrary direction perpendicular to the $Z$ axis. The direction of scattered photon is denoted as $\boldsymbol{k}_{f}$, and the vectors $\boldsymbol{q}$ is defined as $\boldsymbol{q} \equiv \boldsymbol{k}_{i}-\boldsymbol{k}_{f}$. The axis $z$ represents the momentum transfer direction.}
\label{coordinate0}
\end{figure}

The integration in Eq.(\ref{doubly differential cross section1}) is over all components of $\boldsymbol{p}_{i}$, namely $p_{x}$, $p_{y}$, $p_{z}$, respectively. However, when one of these components, such as $p_{z}$, is integrated out, the Dirac delta function $\delta (E_{i}+\omega_{i}-E_{f}-\omega_{f})$ in the integrand constrains $p_{z}$ component to be a fixed value, leaving $p_{x}$ and $p_{y}$ components in the integral. Further, the fixed value for $p_{z}$ component can be completely determined by energy and momentum conservations. For convenience, we can introduce a coordinate system $xyz$ such that $z$ axis represents the momentum transfer direction. In this coordinate system, the $p_{z}$ component can be expressed as:
\begin{equation}
p_{z} = -\frac{\boldsymbol{p}_{i}\cdot\boldsymbol{q}}{q}
      = \frac{\omega_{i}\omega_{f}(1-\cos\theta)-E(p_{z})(\omega_{i}-\omega_{f})}{c^{2}q} \label{projection momentum}
\end{equation}
where $E(p_{z})=\sqrt{m^{2}c^{4}+p_{z}^{2}c^{2}}$ and $q$ is the modulus of the momentum transfer vector $\boldsymbol{q} \equiv \boldsymbol{k}_{i}-\boldsymbol{k}_{f}$. Furthermore, previous study revealed that $p_{z}$ and $E(p_{z})$ are exactly the energy and momentum minimum of the initial state electrons activated in Compton scattering \cite{LaJohn}, namely
\begin{equation}
p_{i}^{\text{min}}=|p_{z}|;\ \ E_{i}^{\text{min}}=E(p_{z}) \label{minimal momentum}
\end{equation}
In many literatures \cite{Ribberfors3,Ribberfors4,Brusa,Salvat,Stutz}, a convenient approximation for $p_{z}$ component is proposed as follows
\begin{equation}
p_{z} \approx \frac{\omega_{i}\omega_{f}(1-\cos\theta)-mc^{2}(\omega_{i}-\omega_{f})}{c^{2}q} \label{projection momentum2}
\end{equation}
This approximation works well in small $p_{z}$ regions, however, it can cause notable discrepancies in large $p_{z}$ regions. The coordinate system $xyz$ is illustrated in Fig. \ref{coordinate0}.

In the previous studies, Ribberfors \emph{et al.} found that the kernel function $X(K_{i},K_{f})$ in Eq. (\ref{doubly differential cross section1}) is a slow-varying function and therefore can be pulled out of the integration \cite{Ribberfors1,Ribberfors3,Brusa}. Successively, this kernel is furthermore approximated by
\begin{eqnarray}
X(K_{i},K_{f}) & \approx & \overline{X}(p_{z}) \nonumber
\\
               &    =    & \frac{K_{i}(p_{z})}{K_{f}(p_{z})}
                           +\frac{K_{f}(p_{z})}{K_{i}(p_{z})} \nonumber
\\
               &         & +2m^{2}c^{2}
                            \bigg(
                              \frac{1}{K_{i}(p_{z})}-\frac{1}{K_{f}(p_{z})}
                            \bigg) \nonumber
\\
               &         & +m^{4}c^{4}
                            \bigg(
                              \frac{1}{K_{i}(p_{z})}-\frac{1}{K_{f}(p_{z})}
                            \bigg)^{2} \label{function X Pmin}
\end{eqnarray}
where
\begin{eqnarray}
K_{i}(p_{z}) & = & \frac{\omega_{i}E(p_{z})}{c^{2}}+\frac{\omega_{i}(\omega_{i}-\omega_{f}\cos\theta)p_{z}}{c^{2}q}
\\
K_{f}(p_{z}) & = & K_{i}(p_{z})-\frac{\omega_{i}\omega_{f}(1-\cos\theta)}{c^{2}}
\end{eqnarray}

Using the above assumptions, the DDCS of Compton scatterings in the former RIA treatments is given by
\begin{eqnarray}
\bigg(\frac{d^{2}\sigma}{d\omega_{f}d\Omega_{f}}\bigg)_{RIA} & = & \frac{r_{0}^{2}}{2}\frac{m}{q}
                                                               \frac{mc^{2}}{E(p_{z})}
                                                               \frac{\omega_{f}}{\omega_{i}}
                                                               \overline{X}(p_{z})
                                                               J(p_{z}) \nonumber
\\
                                                         & = & \overline{Y}^{RIA} \cdot J(p_{z}) \label{RIA}
\end{eqnarray}
where $J(p_{z})$ is an integral for $p_{x}$ and $p_{y}$ components. The same results can be derived from Eq. (\ref{doubly differential cross section1}) through integration by part \cite{Ribberfors1}.

An alternative and simpler approximation of kernel function $X(K_{i},K_{f})$ can be made by taking the $p_{z}\rightarrow0$ limit of $\overline{X}(p_{z})$, which finally gives its Klein-Nishina value \cite{Ribberfors3,Ribberfors4}
\begin{equation}
X(K_{i},K_{f}) \approx X_{KN}=\frac{\omega_{i}}{\omega_{f}}+\frac{\omega_{f}}{\omega_{i}}-\sin^{2}\theta
\end{equation}
Therefore the simplified results of DDCS for Compton scatterings in former RIA treatments can be expressed as
\begin{equation}
\bigg(\frac{d^{2}\sigma}{d\omega_{f}d\Omega_{f}}\bigg)_{RIA}=\frac{r_{0}^{2}}{2}
                                                         \frac{m}{q}\frac{\omega_{f}}{\omega_{i}}
                                                         X_{KN}
                                                         J(p_{z})
                                                        =Y_{KN}^{RIA} \cdot J(p_{z}) \label{RIA simplified}
\end{equation}

From Eq. (\ref{RIA}) and Eq. (\ref{RIA simplified}), it is obvious that the DDCS of Compton scattering in former RIA treatments factorizes into two parts similar to Eq. (\ref{IA}),
\begin{equation}
\bigg(\frac{d^{2}\sigma}{d\omega_{f}d\Omega_{f}}\bigg)_{RIA}=Y^{RIA} \cdot J(p_{z})
\label{RIA factorization}
\end{equation}
The correction factor $J(p_{z})$, which incorporate ground state electron momentum distribution, is called the atomic Compton profile
\begin{equation}
J(p_{z})\equiv\iint\rho(\boldsymbol{p})dp_{x}dp_{y} \label{Compton profile}
\end{equation}
For most of the atomic systems, the momentum distribution is spherical symmetric, then atomic Compton profile reduces to
\begin{equation}\label{electron profile2}
J(p_{z})=2\pi\int\limits_{|p_{z}|}^{\infty}p\rho(p)dp
\end{equation}
In these cases, the Compton profile $J(p_{z})$ is bell-shaped and axisymmetric around the $p_{z}=0$ axis. We restrict ourselves to the spherical symmetric case in this study.

From the procedures described above, we can notice that there are similarities among several former RIA treatments for Compton scattering. The kernel function approximations $X(K_{i},K_{f}) \approx X_{KN}$ and $X(K_{i},K_{f}) \approx \overline{X}(p_{z})$, together with the factorization result Eq. (\ref{RIA factorization}), are the key features for former RIA treatments. Currently, these former RIA treatments are still directly used in the theoretical and simulative studies \cite{Brusa,Salvat,Stutz}. Moreover, large numbers of interdisciplinary works in condensed matter physics and material science focusing on the electron correlations \cite{Kubo,Pisani}, electron momentum distributions \cite{Cooper0,Cooper,Aguiar,Gillet}, band structures, and Fermi surfaces \cite{Wang,Rathor}, are based on these approximations. In the next subsection, we introduce an ``exact'' numerical approach to calculate the DDCS of Compton scattering, which do not utilise the above kernel function approximations and the factorization results. Therefore, in principle, our approach is more precise than those of former RIA treatments. Furthermore, the validity of the above kernel function approximations and the factorization results, which have been widely adopted in former RIA treatments as well as interdisciplinary studies, can be rigorously tested using our ``exact'' numerical approach.

\subsection{``Exact'' Numerical Treatments \label{sec:2b}}

\begin{figure}
\includegraphics[width=0.565\textwidth]{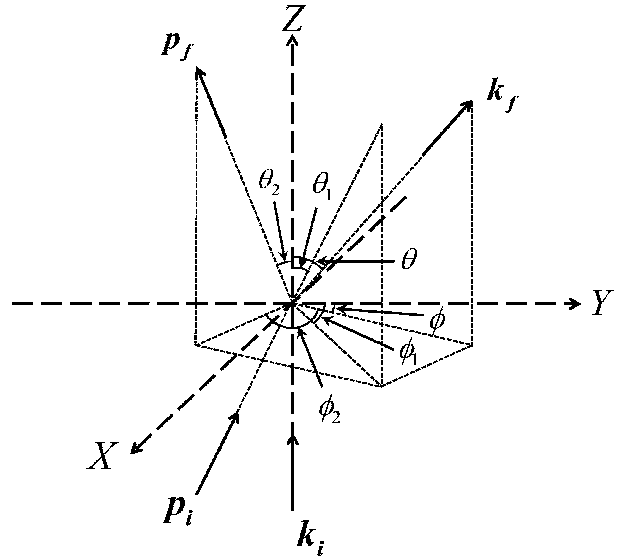}
\caption{Coordinate system $XYZ$ used in the numerical evaluation. The $Z$ axis is chosen to be the direction of initial photon $\boldsymbol{k}_{i}$ similar to Fig. \ref{coordinate0}. The direction of scattered photon is denoted as $\boldsymbol{k}_{f}=(k_{f},\theta,\phi)$, and the vectors $\boldsymbol{p}_{i}=(p_{i},\theta_{1},\phi_{1})$ and $\boldsymbol{p}_{f}=(p_{f},\theta_{2},\phi_{2})$ represent the momentum of initial state and final state electron.}
\label{coordinate}
\end{figure}

In this section, we describe our ``exact'' numerical treatment for RIA formulation. Instead of treating the kernel function $X(K_{i},K_{f})$ to be a slow-varying function in the integration as Ribberfors \emph{et al.} expected, we directly evaluate the integral in Eq. (\ref{doubly differential cross section1}) through numerical scheme.

The geometry of Compton scattering process efficient to numerical evaluation is illustrated in Fig. \ref{coordinate}. We choose a coordinate system $XYZ$ such that the incoming photon move towards the $Z$ direction, the azimuthal angle and polar angle for outgoing photon, initial electron, and final electron are denoted as $(\theta,\phi)$, $(\theta_{1},\phi_{1})$, $(\theta_{2},\phi_{2})$, respectively. By employing such coordinate system, the function $K_{i}$ and $K_{f}$ can be calculated as
\begin{eqnarray}
K_{i} & = & K_{i}(p_{i},\theta_{1})
        =   \frac{E_{i}\omega_{i}}{c^{2}}
            -\frac{\omega_{i}p_{i}\cos{\theta_{1}}}{c} \label{1}
\\
K_{f} & = & K_{f}(p_{i},\theta_{1})
        =  K_{i}(p_{i},\theta_{1})-\frac{\omega_{i}\omega_{f}(1-\cos\theta)}{c^{2}} \label{2}
\end{eqnarray}
From the energy and momentum conservations in Compton scattering process, the energy of scattered electron is given by
\begin{eqnarray}
E_{f} & = & E_{f}(p_{i},\theta_{1},\phi_{1}) \nonumber
\\
      & = & \sqrt{p_{i}^{2}c^{2}+m^{2}c^{4}+\omega_{i}^{2}+\omega_{f}^{2}-2\omega_{i}\omega_{f}\cos{\theta}} \nonumber
\\
      &   & \overline{-2p_{i}c\omega_{f}[\cos{\theta}\cos{\theta_{1}}+\sin{\theta}\sin{\theta_{1}}\cos(\phi-\phi_{1})]} \nonumber
\\
      &   & \overline{+2p_{i}c\omega_{i}\cos{\theta_{1}}} \label{final-electron-energy}
\end{eqnarray}

For simplicity, in this work, we only restrict ourselves to the spherical symmetric atomic systems, and more complicated molecular or condense matter systems are not taken into consideration. In our numerical calculations, Dirac-Fock theory is used to achieve the ground state wavefunctions and electron's momentum distribution. Since spherical symmetric atomic systems are considered, the electron's momentum distribution reduces to $\rho(\boldsymbol{p}_{i})=\rho(p_{i})$. In the Dirac-Fock theory, the wavefunction of an individual electron is given by the Dirac orbital $u_{njl}(r)$, which is composed of a large component $G_{njl}$ and a small component $F_{njl}$. Then the corresponding large and small components of momentum wavefunctions are given by the following Fourier transformation:
\begin{eqnarray}
\phi_{njl}^{G}(p) & = & \sqrt{\frac{2}{\pi}} \int_{0}^{\infty} G_{nlj}(r)j_{l}(pr)r^{2}dr \\
\phi_{njl}^{F}(p) & = & \bigg\{
                          \begin{array}{cc}
                           \sqrt{\frac{2}{\pi}} \int_{0}^{\infty}F_{njl}(r)j_{l+1}(pr)r^{2}dr & j=l+\frac{1}{2} \\
                           \sqrt{\frac{2}{\pi}} \int_{0}^{\infty}F_{njl}(r)j_{l-1}(pr)r^{2}dr & j=l-\frac{1}{2}
                          \end{array} \nonumber
\\
\end{eqnarray}
and the total momentum distribution can be calculated through
\begin{eqnarray}
\rho(p_{i}) & = & \sum_{a}|\phi_{a}(p_{i})|^{2} \nonumber
\\
            & = & \sum_{njl}N_{njl}
                  \bigg(
                    (\phi_{njl}^{G}(p_{i}))^{2}+(\phi_{njl}^{F}(p_{i}))^{2}
                  \bigg) \label{rho}
\end{eqnarray}
where $N_{njl}$ is the number of electrons in each orbital $(njl)$. The detailed descriptions on the Dirac orbital $u_{njl}(r)$ as well as its large and small components are given in Appendix \ref{appendix0}.

Put the Eq (\ref{1}), Eq. (\ref{2}) and Eq. (\ref{rho}) into the integration in Eq. (\ref{doubly differential cross section1}), and take atomic binding energies into account, we obtain the DDCS for Compton scattering processes
\begin{eqnarray}
\frac{d^{2}\sigma}{d\omega_{f}d\Omega_{f}} & = & \sum_{njl}\frac{d^{2}\sigma_{njl}}{d\omega_{f}d\Omega_{f}} \nonumber
\\
                                           & = & \sum_{njl}\frac{r_{0}^{2}}{2} \frac{\omega_{f}}{\omega_{i}} m^{2}c^{4}
                                                 \Theta(\omega_{i}-\omega_{f}-E_{njl}^{B})N_{njl} \nonumber
\\
                                           &   & \iiint{p_{i}^{2} dp_{i} \sin{\theta_{1}} d\theta_{1} d\phi_{1}}
                                                 \delta (E_{i}+\omega_{i}-E_{f}-\omega_{f}) \nonumber
\\
                                           &   & \ \ \ \ \ \times
                                                 \bigg(
                                                   (\phi_{njl}^{G}(p_{i}))^{2}+(\phi_{njl}^{F}(p_{i}))^{2}
                                                 \bigg) \nonumber
\\
                                           &   & \ \ \ \ \ \times
                                                 \frac{X(K_{i}(p_{i},\theta_{1}),K_{f}(p_{i},\theta_{1}))}{E_{i}(p_{i})E_{f}(p_{i},\theta_{1},\phi_{1})}
\label{doubly differential cross section2}
\end{eqnarray}
where $E_{njl}^{B}$ is the binding energy of orbital ($njl$), and $\Theta(\omega_{i}-\omega_{f}-E_{njl}^{B})$ is the Heaviside step function. When the energy transfer $T=\omega_{i}-\omega_{f}$ is less than atomic binding energy $E_{njl}^{B}$, the Heaviside step function vanishes cross section from this orbital $(njl)$. In other words, electron in this orbital is inactive in atomic Compton scattering process $\hbar\omega_{i}+ A \longrightarrow \hbar\omega_{f}+e^{-}+ A^{+}$.

In order to get the results of DDCS numerically, one point should be mentioned. In Eq. (\ref{doubly differential cross section2}), when integrating one of the three variables $p_{i}$, $\theta_{1}$ and $\phi_{1}$, the Diarc delta function $\delta(E_{i}+\omega_{i}-E_{f}-\omega_{f})$ in the integrand restricts this variable to a fixed value. The fixed values $\widetilde{p}_{i}$, $\widetilde{\theta}_{1}$ or $\widetilde{\phi}_{1}$ can be solved by finding the zeros of function
\begin{equation}
f(p_{i},\theta_{1},\phi_{1})=E_{i}(p_{i})+\omega_{i}-E_{f}(p_{i},\theta_{1},\phi_{1})-\omega_{f} \label{function f}
\end{equation}
where $E_{i}(p_{i})=\sqrt{p_{i}^{2}c^{2}+m^{2}c^{4}}$ and $E_{f}(p_{i},\theta_{1},\phi_{1})$ is calculated in Eq. (\ref{final-electron-energy})

To evaluate the integral in Eq. (\ref{doubly differential cross section2}), we first integrate over the azimuthal angle $\phi_{1}$. After some redundant calculations routinely, we get the DDCS for Compton scatterings:
\begin{eqnarray}
\frac{d^{2}\sigma}{d\omega_{f}d\Omega_{f}} & = & \sum_{njl}
                                                 \frac{r_{0}^{2}}{2\omega_{i}\sin{\theta}} m^{2}c^{4}
                                                 \Theta(\omega_{i}-\omega_{f}-E_{njl}^{B})N_{njl} \nonumber
\\
                                           &   & \iint{p_{i} dp_{i} d\theta_{1}}
                                                 \bigg(
                                                   (\phi_{njl}^{G}(p_{i}))^{2}+(\phi_{njl}^{F}(p_{i}))^{2}
                                                 \bigg) \nonumber
\\
                                           &   & \times \frac{X(K_{i}(p_{i},\theta_{1}),K_{f}(p_{i},\theta_{1}))}
                                                             {E_{i}(p_{i})\times c\sqrt{1-\cos^{2}(\phi-\widetilde{\phi}_{1})}}
\label{doubly differential cross section4}
\end{eqnarray}
where the fixed azimuthal angle $\widetilde{\phi}_{1}$ satisfies
\begin{eqnarray}
\cos(\phi-\widetilde{\phi}_{1}) & = & \frac{\omega_{i}^{2}+\omega_{f}^{2}-2\omega_{i}\omega_{f}\cos{\theta}}{2p_{i}c\omega_{f}\sin{\theta}\sin{\theta_{1}}}
                                        -\frac{(\omega_{i}-\omega_{f})^{2}}{2p_{i}c\omega_{f}\sin{\theta}\sin{\theta_{1}}} \nonumber
\\
                                  &   & -\frac{(\omega_{i}-\omega_{f})E_{i}(p_{i})}
                                              {p_{i}c\omega_{f}\sin{\theta}\sin{\theta_{1}}}
                                        +\frac{\cos{\theta_{1}}(\omega_{i}-\omega_{f}\cos{\theta})}{\omega_{f}\sin{\theta}\sin{\theta_{1}}} \nonumber
\\
\label{fix-gamma}
\end{eqnarray}
Moreover, it is worth noting that, only those which satisfy Eq. (\ref{fix-gamma}) and the inequality $-1 \leq \cos(\phi-\widetilde{\phi}_{1}) \leq 1$ simultaneously can be regarded as physical allowed value of $\widetilde{\phi}_{1}$.

In this work, we adopt the aforementioned order of integration in the numerical evaluation of Eq. (\ref{doubly differential cross section2}). However, equivalent results can be achieved by exchanging the order of integration. Results obtained from alternative order of integration are given in Appendix \ref{appendix1}.

\section{Results and Discussions\label{sec:3}}

\begin{figure*}
\centering
\includegraphics[width=0.495\textwidth]{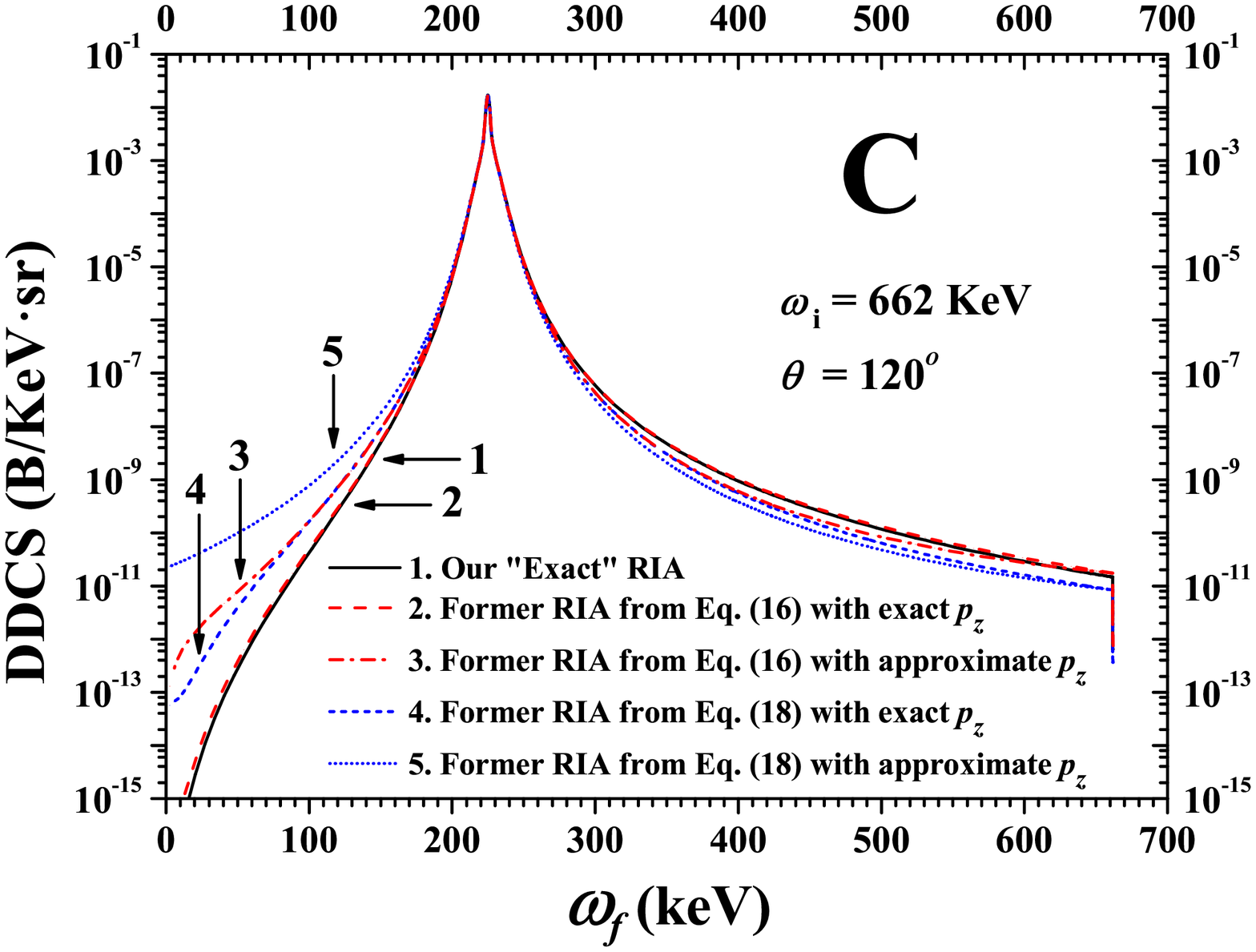}
\includegraphics[width=0.495\textwidth]{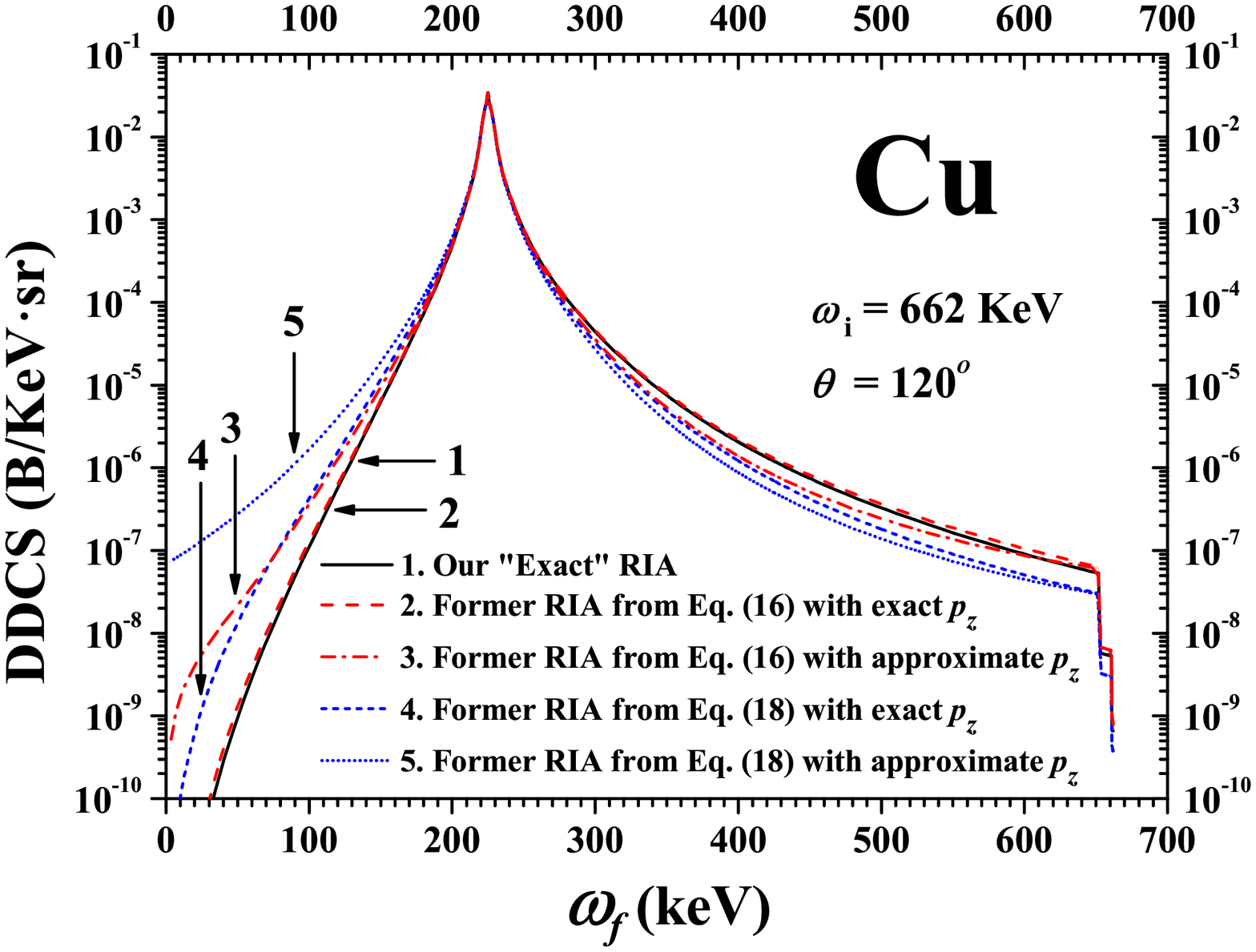}
\includegraphics[width=0.495\textwidth]{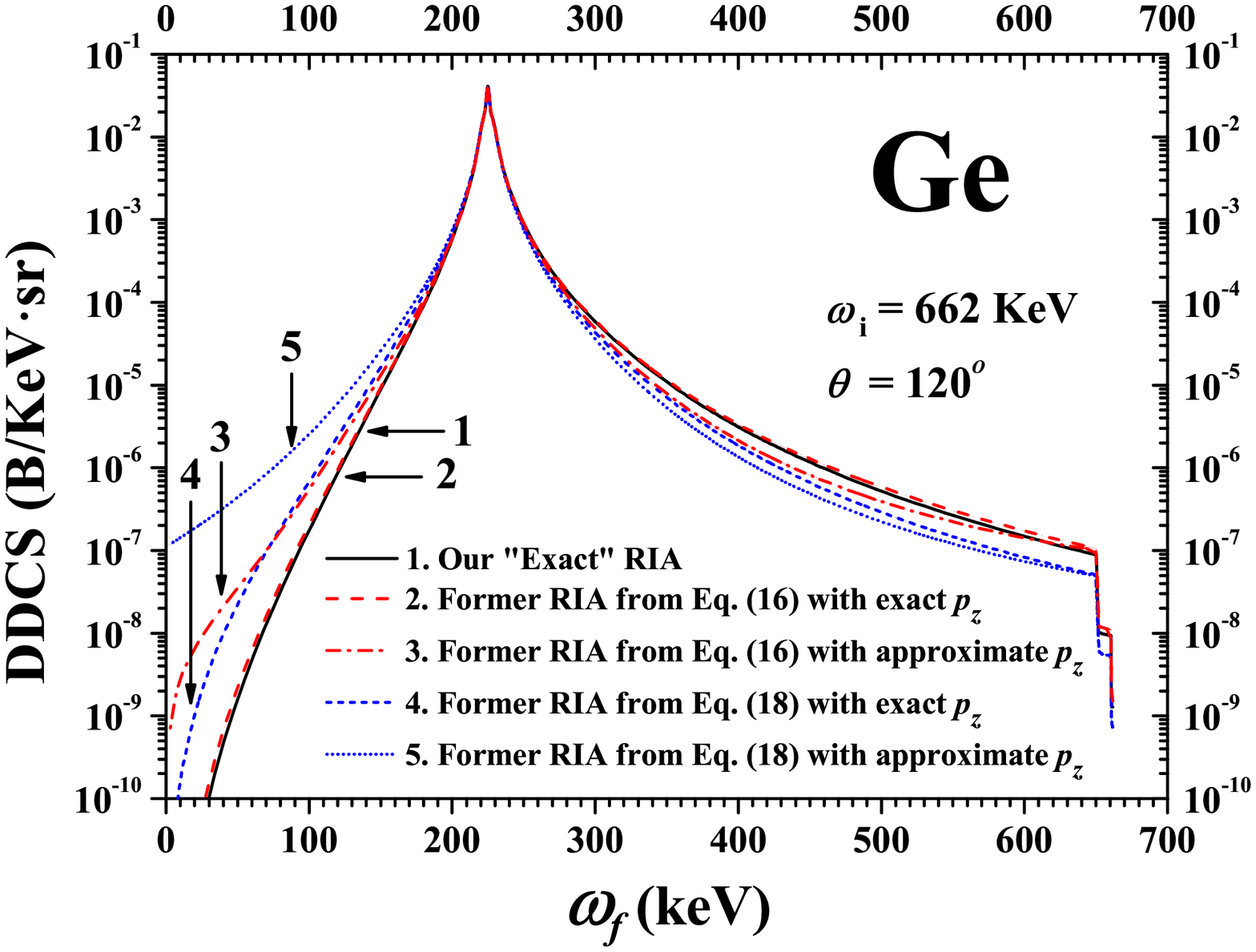}
\includegraphics[width=0.495\textwidth]{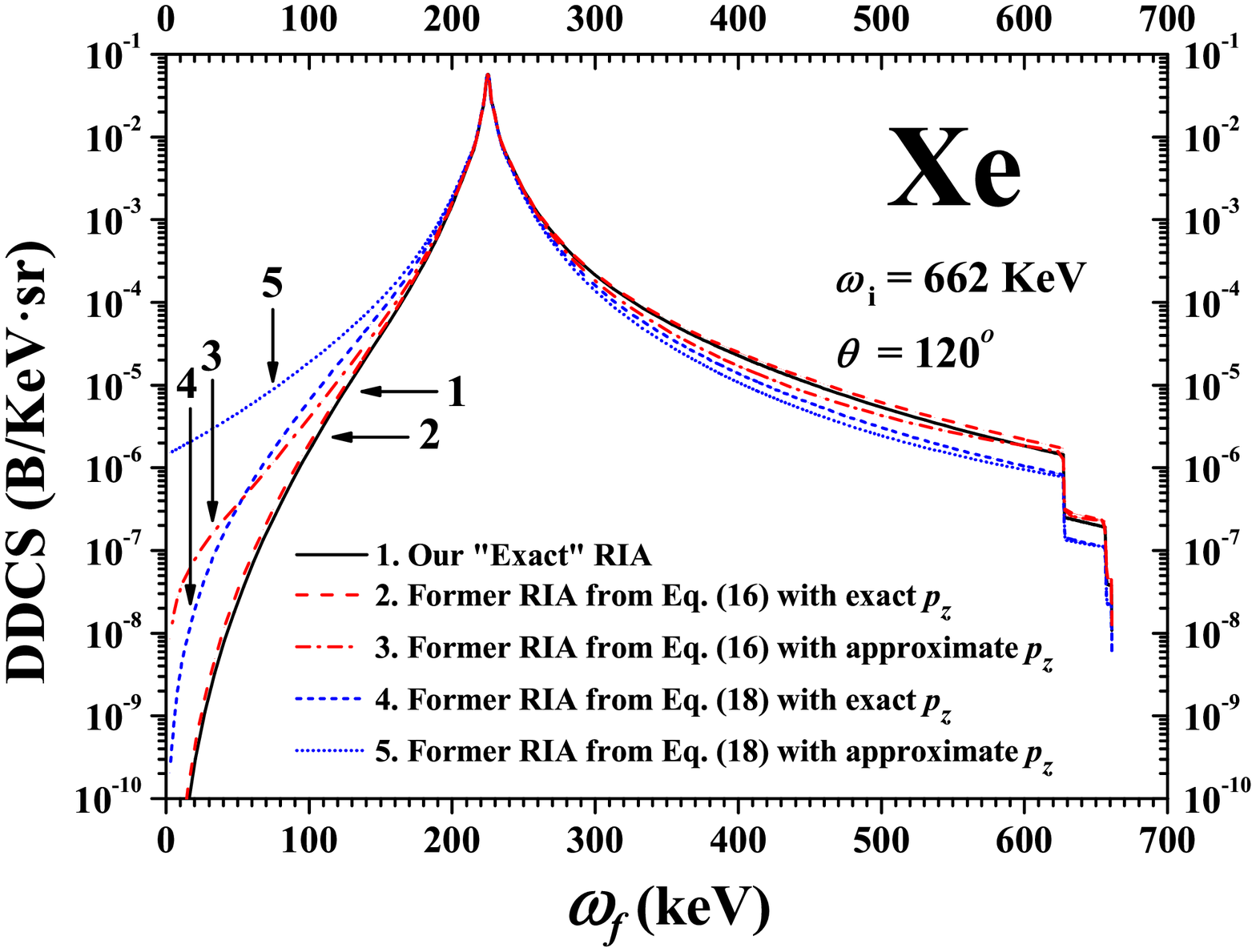}
\caption{DDCS of Compton scattering obtained for C, Cu, Ge and Xe atoms at photon energy $\omega_{i}=662$ KeV and scattering angle $\theta=120^{\text{o}}$. The results of our method and those of several former treatments of RIA are shown. The curves in this figure represent: 1) solid lines -- the results of our ``exact'' RIA treatment; 2) dashed lines -- the results of former RIA treatment employing Eq. (\ref{RIA}) with exact $p_{z}$ values calculated in Eq. (\ref{projection momentum}); 3) dashed-dotted lines -- the results of former RIA treatment utilizing Eq. (\ref{RIA}) with approximate $p_{z}$ values computed in Eq. (\ref{projection momentum2}); 4) short-dashed lines -- the results of former RIA treatment using Eq. (\ref{RIA simplified}) with exact $p_{z}$ values; 5) short-dotted curves -- the results of former RIA treatment using Eq. (\ref{RIA simplified}) with approximate $p_{z}$ values.}
\label{DDCS_figure1}
\end{figure*}

In this section, we provide the results of atomic Compton scattering obtained using our ``exact'' numerical method of RIA described in Section \ref{sec:2b}. For a comprehensive study, we choose four neutral atoms C, Cu, Ge, and Xe to represent the small-$Z$, middle-$Z$, and large-$Z$ regimes. Section \ref{sec:3a} is focused on differential cross sections, where a detailed comparison of our results and those of former treatments of RIA is presented. The validity of the factorization in Eq. (\ref{IA}) and the available ranges of former RIA treatments are discussed using this comparison. In Section \ref{sec:3b}, effective Compton profiles are extracted from our results to quantitatively illustrate the differences between our method and the former RIA treatments. Furthermore, we provide an uncertainty estimate for the numerical scheme in Section \ref{sec:3c}.

\subsection{Differential Cross Sections \label{sec:3a}}

\begin{figure*}
\centering
\includegraphics[width=0.495\textwidth]{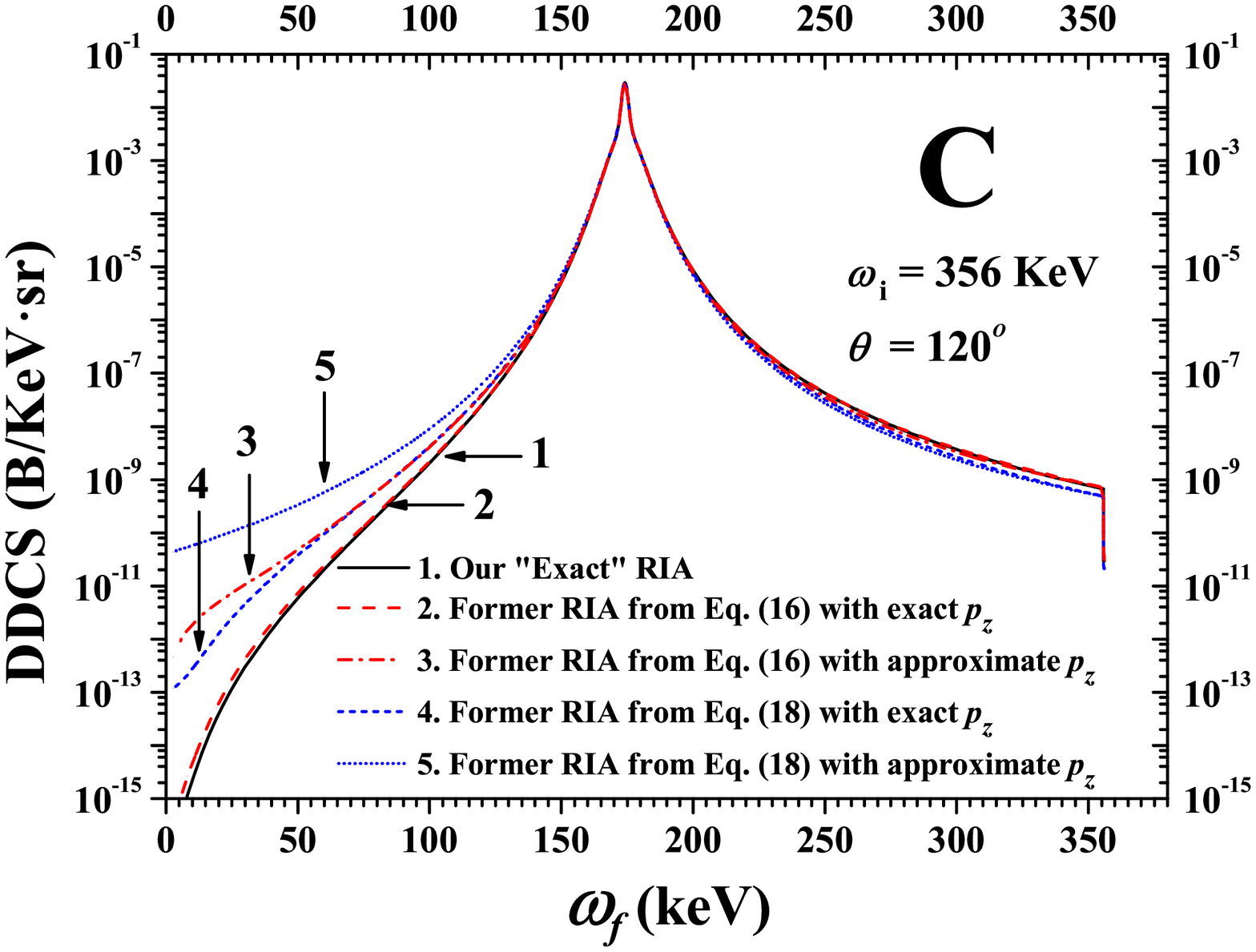}
\includegraphics[width=0.495\textwidth]{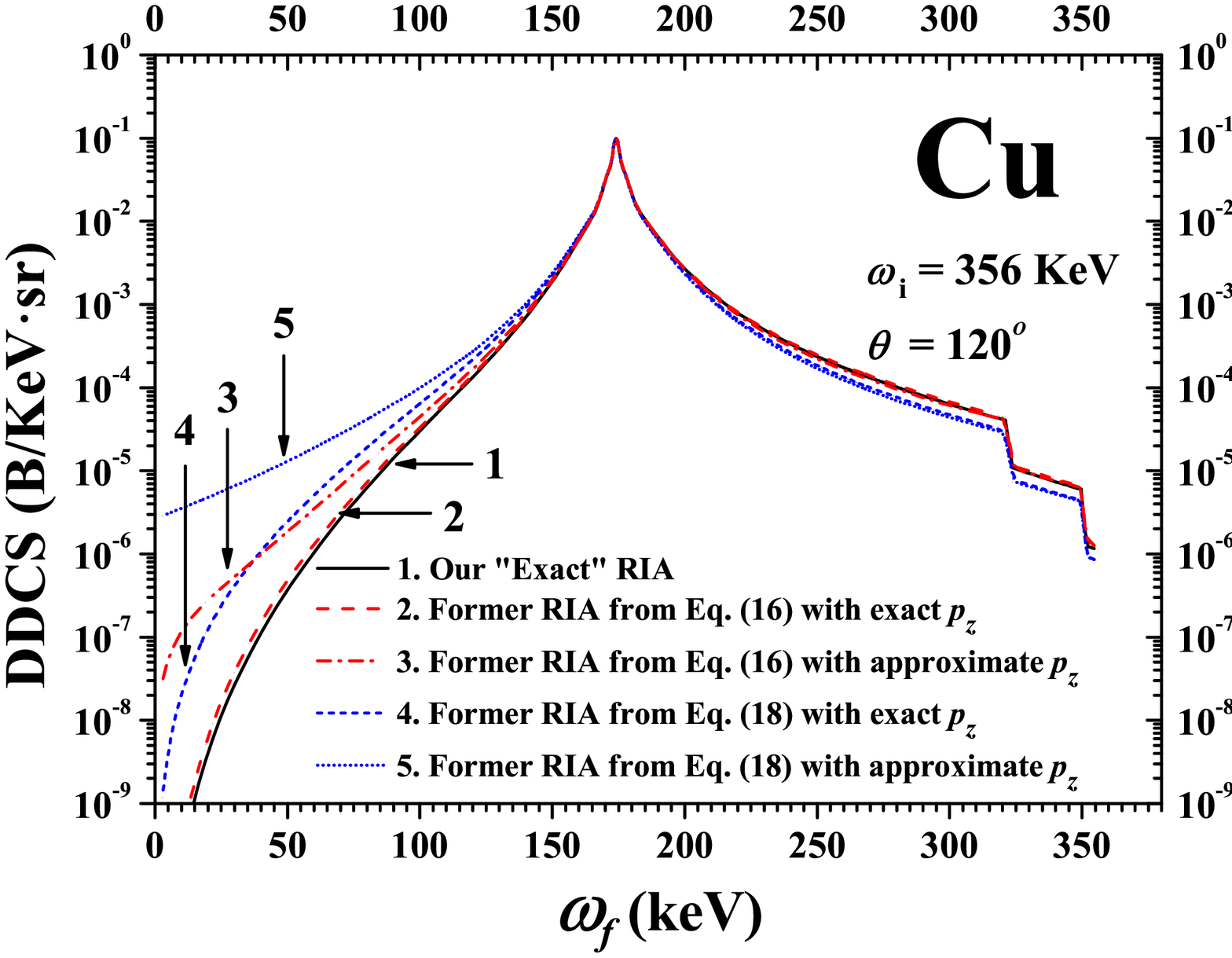}
\includegraphics[width=0.495\textwidth]{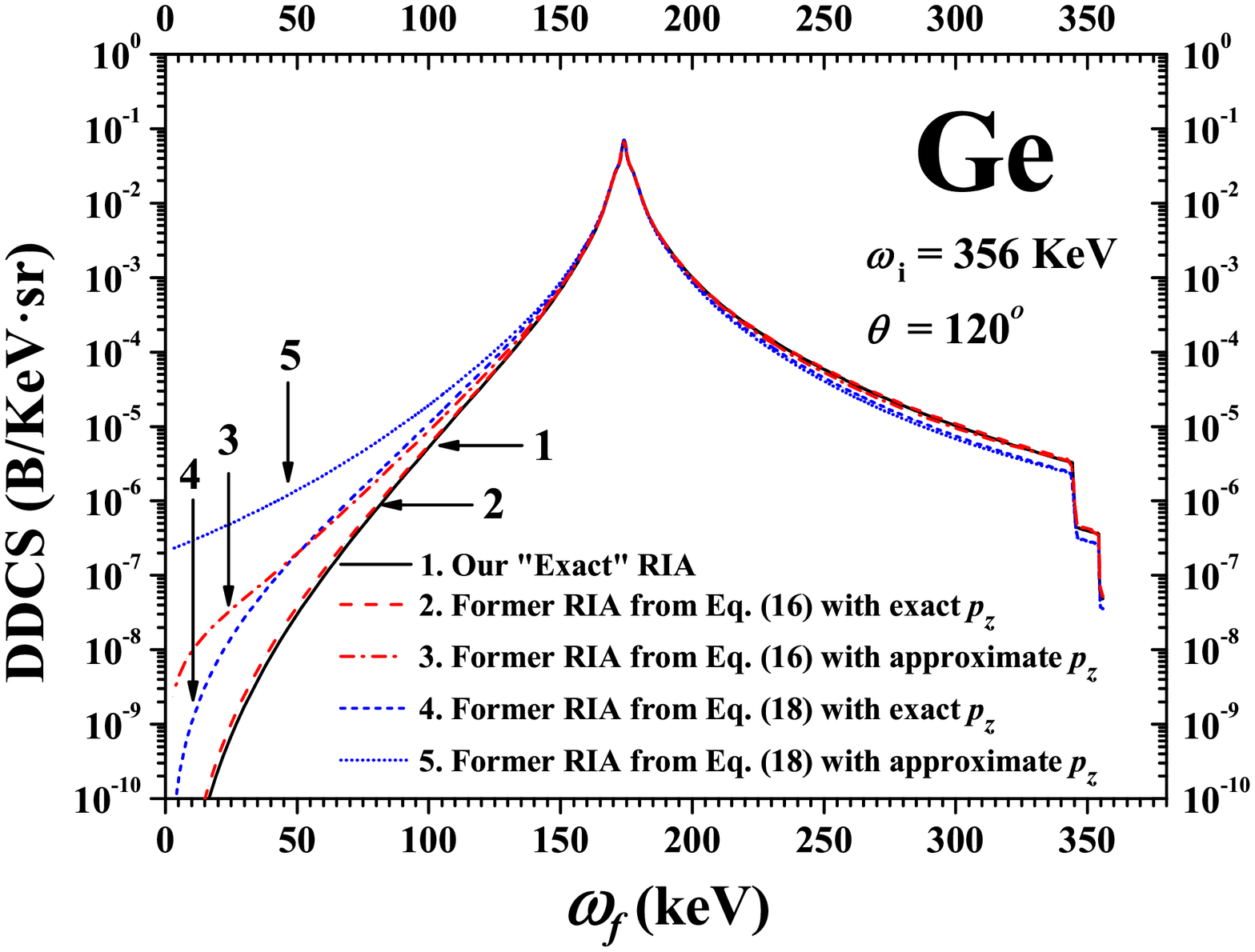}
\includegraphics[width=0.495\textwidth]{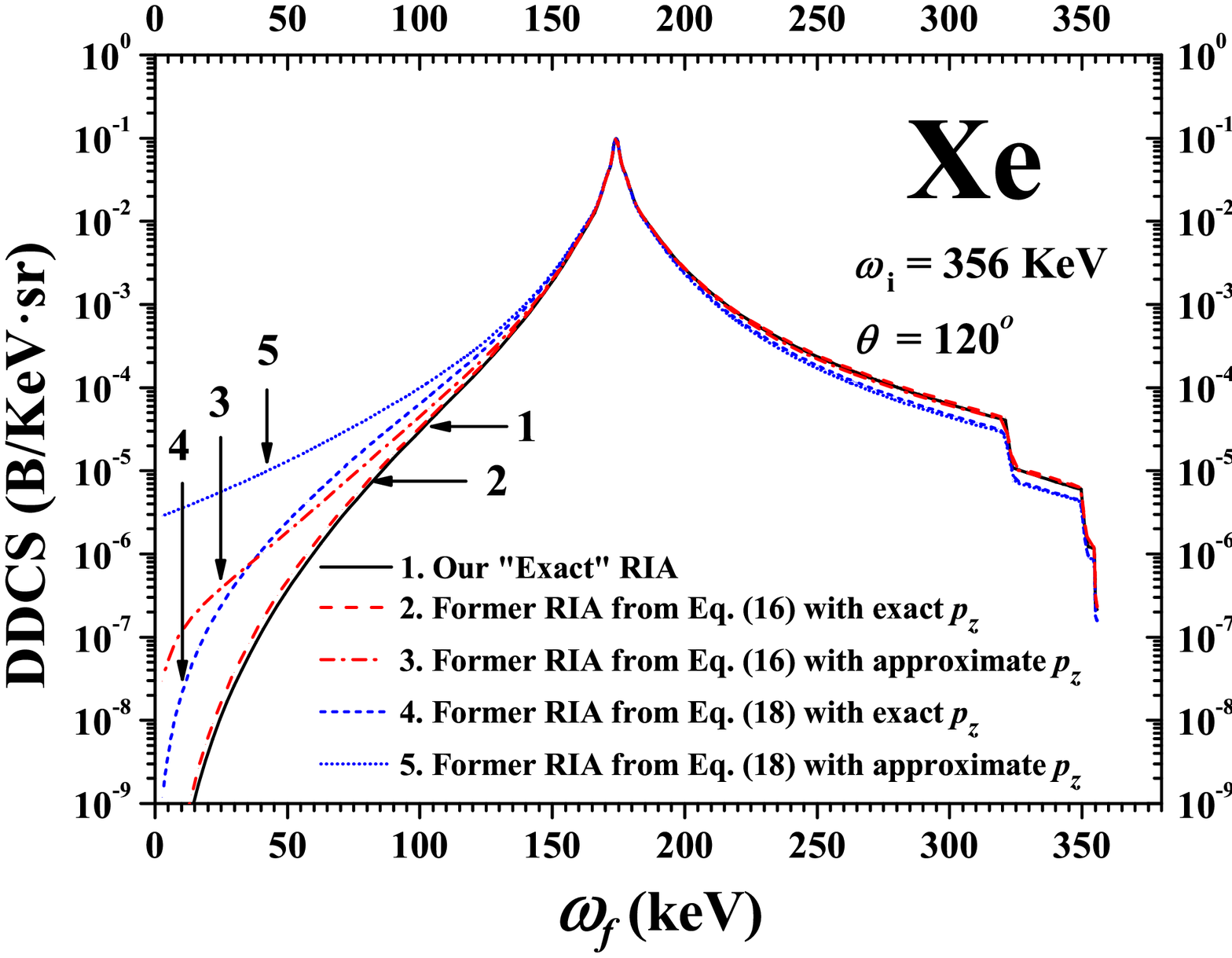}
\caption{DDCS of Compton scattering obtained for C, Cu, Ge and Xe atoms at photon energy $\omega_{i}=356$ KeV and scattering angle $\theta=120^{\text{o}}$. The results of our ``exact'' RIA treatment and several former RIA treatments are plotted similar to that in Fig. \ref{DDCS_figure1}.}
\label{DDCS_figure2}
\end{figure*}

In this subsection, we focus on the differential cross sections in atomic Compton scatterings. The DDCS of Compton Scatterings for C, Cu, Ge, and Xe atoms at photon energies $\omega_{i}=662$ KeV, $356$ KeV and a scattering angle $\theta=120^{\text{o}}$ are shown in Fig. \ref{DDCS_figure1} and Fig. \ref{DDCS_figure2}. Comparative results of our ``exact'' numerical treatment of RIA and several former treatments of RIA have been illustrated in this figure. The results of former RIA treatments are obtained using Eq. (\ref{RIA}) and Eq. (\ref{RIA simplified}), where the DDCS of Compton scatterings is factorized into $Y$ times atomic Compton profiles $J$, similar to Eq. (\ref{IA}). Moreover, when computing the atomic Compton profiles, the momentum component $p_{z}$ can be calculated using its exact or approximate values obtained from Eq. (\ref{projection momentum}) and Eq. (\ref{projection momentum2}). Our ``exact'' RIA results are obtained by directly evaluating the numerical integral in Eq. (\ref{doubly differential cross section1}).

The numerical results in Fig. \ref{DDCS_figure1} and Fig. \ref{DDCS_figure2} show that the DDCS of Compton scattering at photon energies $662$ KeV and $356$ KeV are very similar, except that the locations of Compton peaks are shifted. The results in these figures indicate that our ``exact'' RIA results confirm the results of former RIA treatments near the Compton peak region $\omega_{f}\approx\omega_{C}$. Thus, we can draw a conclusion that the factorization treatments adopted in former RIA studies do not change the physical results significantly in the Compton peak region, as Ribberfors \emph{et al.} expected \cite{Ribberfors1,Ribberfors3,Ribberfors4}. Furthermore, recent works have indicated that the available range of former RIA treatments is only near the Compton peak region, where the momentum component $|p_{z}|$ in the scattering process is not very large \cite{Pratt,Kaliman,Pratt3,Pratt2,Drukarev0}. This criterion can be demonstrated by comparison with more advanced approaches, such as the S-Matrix approach. Therefore, in the region where the RIA formulation is believed to be valid, the factorization treatments, as well as kernel function approximations, used in former RIA studies still hold, and produce only a few deviations in the integration of DDCS. Since most of interdisciplinary studies in condensed matter physics and material science using former RIA treatments consider only the cases near the Compton peak region, they are reliable at a sufficient accuracy \cite{Kubo,Pisani,Cooper0,Cooper,Aguiar,Wang,Gillet,Sahariya,Rathor}. In Appendix \ref{appendix5}, we have provided comparative results on the DDCS of Compton scattering between RIA and S-Matrix approaches.

However, when the energy of scattered photon is far from the Compton peak region, discrepancies between our ``exact'' RIA treatment and several former RIA treatments become notable. Therefore, some factorization treatments and kernel function approximations used in former RIA studies are invalid in this region; in addition, they produce non-negligible deviations in the integration of DDCS. Moreover, when $\omega_{f}<\omega_{C}$, former RIA treatments overestimate the DDCS of Compton scattering, whereas in the region $\omega_{f}>\omega_{C}$, our results obtain larger cross sections than the former RIA results. In several former RIA treatments, both the approximations of kernel function $X(K_{i},K_{f})$ and the values of momentum component $p_{z}$ significantly impact the DDCS of Compton scattering. Among the former treatments of RIA, only one approach utilized more accurate kernel function approximation $X(K_{i},K_{f})\approx \overline{X}(p_{z})$ with the exact momentum component $p_{z}$ values, which correspond to dashed curves in Fig. \ref{DDCS_figure1} and Fig. \ref{DDCS_figure2}, agrees well with our approach in the entire energy spectrum. Further, in the Appendix \ref{appendix5}, through the comparisons with theoretical S-Matrix calculations and experimental results, we can observe that the available range of our ``exact'' RIA treatment is still only near the Compton peak, which is similar to that in former RIA treatments. In regions far from the Compton peak, our approach, despite employing an exact scheme in the numerical integration, does not exhibit a significant improvement over the former RIA treatments.

To directly validate the former RIA treatments using the kernel function, we numerically study the function $X(K_{i},K_{f})$ in Eq. (\ref{function X}), and the results are provided in Appendix \ref{appendix}. A detailed analysis indicates that the kernel approximation $X(K_{i},K_{f})\approx X_{KN}$ in former RIA treatments works well only in the Compton peak region $\omega_{f}\approx\omega_{C}$. When the final photon energy reaches far beyond the Compton peak region, the kernel function approximation $X(K_{i},K_{f})\approx X_{KN}$ becomes inappropriate. The result is consistent with our conclusion obtained from the DDCS of Compton scattering in Fig. \ref{DDCS_figure1} and Fig. \ref{DDCS_figure2}, where the results of former RIA treatments based on the approximation $X(K_{i},K_{f})\approx X_{KN}$ through Eq. (\ref{RIA simplified}) exhibit notable discrepancies when final photon energy is far from the Compton peak region, irrespective of the employment of exact or approximate $p_{z}$ values. For comparison, in Appendix \ref{appendix}, we demonstrate that $X(K_{i},K_{f})\approx \overline{X}(p_{z})$ is a more accurate kernel function approximation than $X(K_{i},K_{f})\approx X_{KN}$ when outgoing photon energy is far from the Compton peak region. Moreover, this conclusion can be revealed from the DDCS of Compton scattering illustrated in Fig. \ref{DDCS_figure1} and Fig. \ref{DDCS_figure2}, where the results of RIA treatments that depend on $X(K_{i},K_{f})\approx \overline{X}(p_{z})$ through Eq. (\ref{RIA}) show lesser discrepancy with our ``exact'' RIA results than those employing $X(K_{i},K_{f})\approx X_{KN}$ through Eq. (\ref{RIA simplified}).

\subsection{Effective Compton Profiles \label{sec:3b}}

\begin{figure*}
\includegraphics[width=0.495\textwidth]{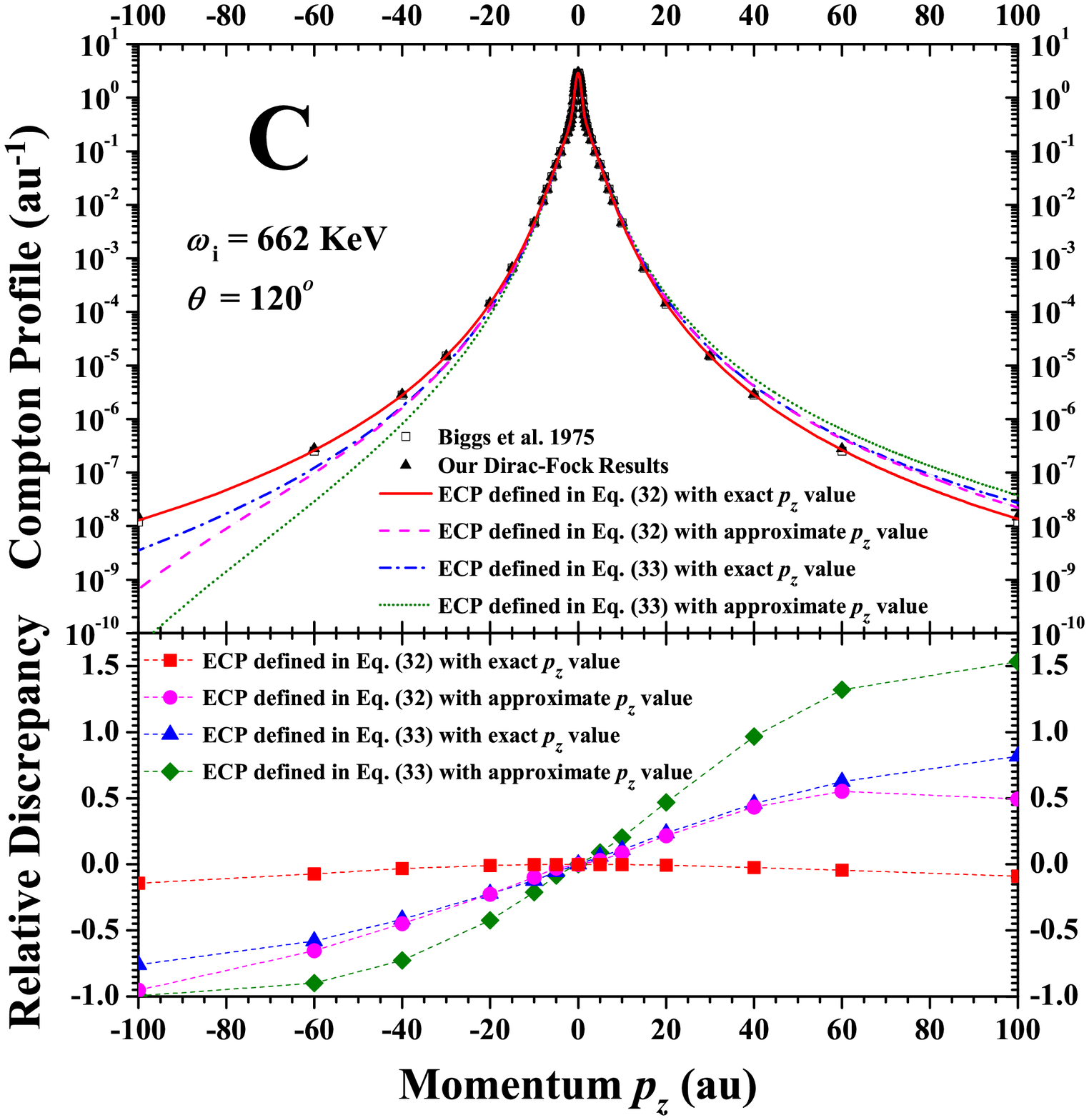}
\includegraphics[width=0.495\textwidth]{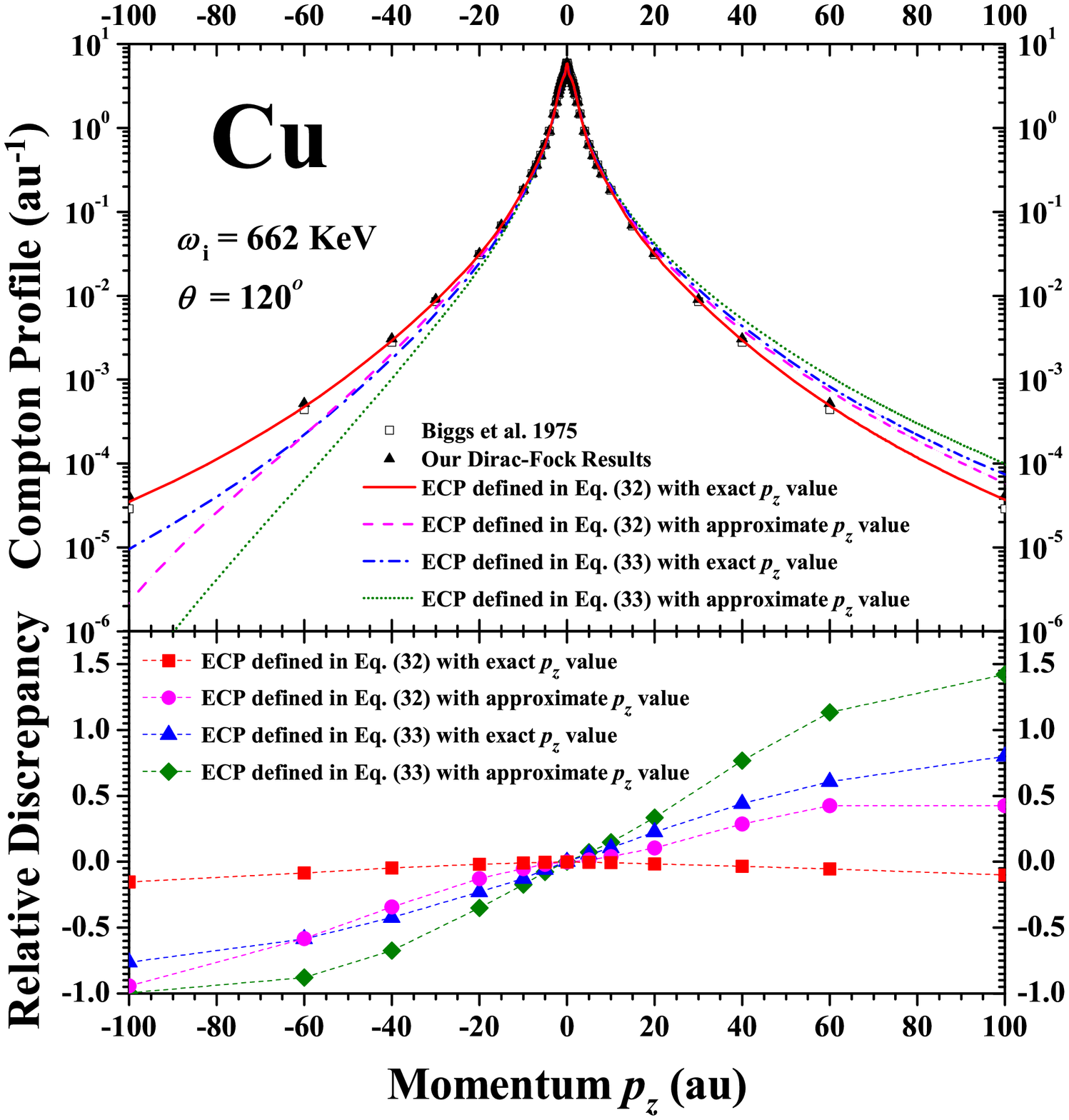}
\includegraphics[width=0.495\textwidth]{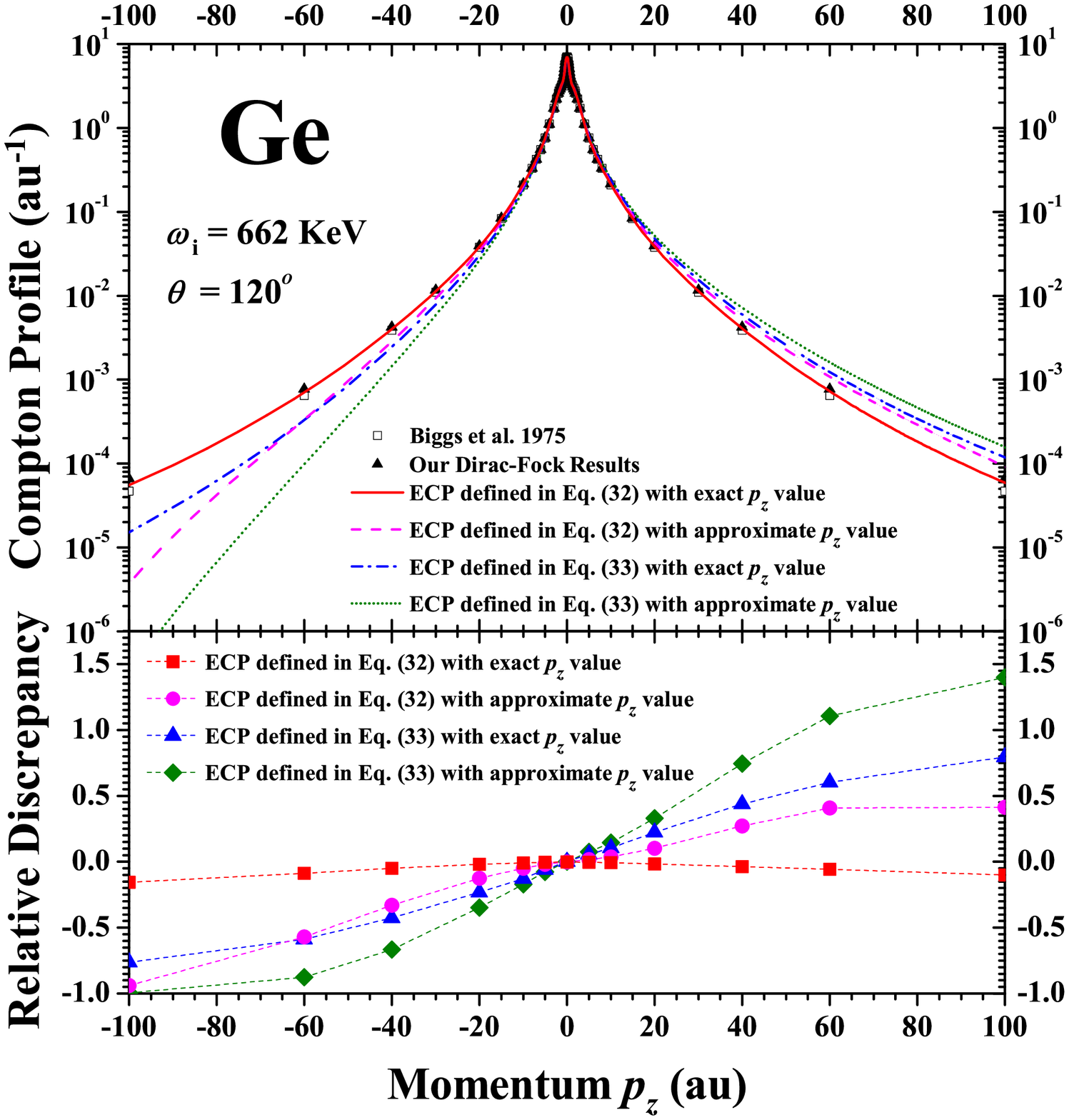}
\includegraphics[width=0.495\textwidth]{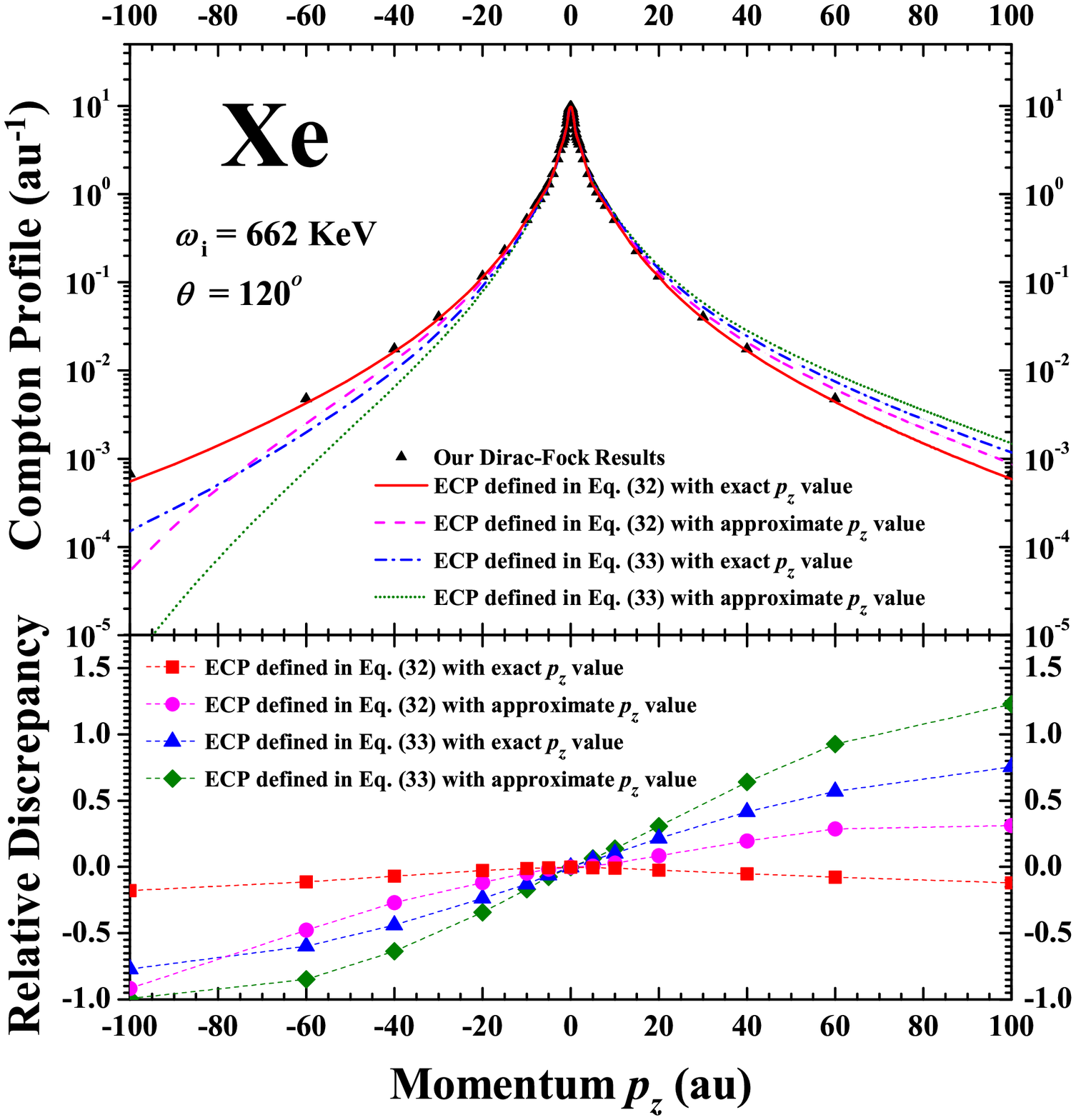}
\caption{Effective Compton profiles (ECP) for C, Cu, Ge, and Xe atoms at a photon energy $\omega_{i}=662$ KeV with a scattering angle $\theta=120^{\text{o}}$. The solid lines correspond to the effective Compton profiles $\overline{J}_{\text{eff}}(p_{z},\omega_{i},\theta)$ defined in Eq. (\ref{effective Compton Profile}) with exact $p_{z}$ values calculated in Eq. (\ref{projection momentum}). The dashed lines correspond to the effective Compton profiles $\overline{J}_{\text{eff}}(p_{z},\omega_{i},\theta)$ defined in Eq. (\ref{effective Compton Profile}) with approximate $p_{z}$ values computed in Eq. (\ref{projection momentum2}). The dashed-dotted lines represent the effective Compton profiles $J_{\text{eff}}(p_{z},\omega_{i},\theta)$ defined in Eq. (\ref{effective Compton Profile1}) with exact $p_{z}$ values calculated in Eq. (\ref{projection momentum}). The short-dotted lines represent the effective Compton profiles $J_{\text{eff}}(p_{z},\omega_{i},\theta)$ defined in Eq. (\ref{effective Compton Profile1}) with approximate $p_{z}$ values computed in Eq. (\ref{projection momentum2}). In addition, the atomic Compton profiles computed using Eq. (\ref{Compton profile}) based on the nonrelativistic Hartree-Fock (HF) theory and the relativistic Dirac-Fock (DF) theory are presented. The HF results are provided by Biggs \emph{et al.} for $Z<36$ \cite{Biggs}, and the DF results are computed using our program. Moreover, the relative discrepancies, which are defined as $D\equiv (J_{\text{eff}}-J)/J$ with $J$ and $J_{\text{eff}}$ as the atomic and effective Compton profiles, are superimposed in the figure for various effective Compton profiles. \label{ECP0}}
\end{figure*}

To further quantitatively compare the results of our method and former RIA treatments quantitatively, we define the effective Compton profiles as
\begin{eqnarray}
\overline{J}_{\text{eff}}(p_{z},\omega_{i},\theta) & \equiv & \frac{1}{\overline{Y}^{RIA}}
                                                              \frac{d^{2}\sigma}{d\omega_{f}d\Omega_{f}} \nonumber
\\
                                                   &   =    & \frac{2}{r_{0}^{2}}\frac{E(p_{z})}{mc^{2}}
                                                              \frac{q}{m}\frac{\omega_{i}}{\omega_{f}}
                                                              \frac{\frac{d^{2}\sigma}{d\omega_{f}d\Omega_{f}}}
                                                                   {\overline{X}(p_{z})} \label{effective Compton Profile}
\end{eqnarray}
and
\begin{eqnarray}
J_{\text{eff}}(p_{z},\omega_{i},\theta) & \equiv & \frac{1}{Y_{KN}^{RIA}}
                                                   \frac{d^{2}\sigma}{d\omega_{f}d\Omega_{f}} \nonumber
\\
                                        &   =    & \frac{2}{r_{0}^{2}}
                                                   \frac{q}{m}\frac{\omega_{f}}{\omega_{i}}
                                                   \frac{\frac{d^{2}\sigma}{d\omega_{f}d\Omega_{f}}}
                                                        {X_{KN}} \label{effective Compton Profile1}
\end{eqnarray}
where $p_{z}$ is the projection of the electron's momentum on the momentum transfer direction, which is calculated by Eq. (\ref{projection momentum}) or Eq. (\ref{projection momentum2}). The differential cross section $d^{2}\sigma/d\omega_{f}d\Omega_{f}$ is numerically obtained using our ``exact'' RIA treatment. If former RIA treatments are adopted in calculating the differential cross sections, the effective Compton profiles automatically reduce to the conventional atomic Compton profile defined in Eq. (\ref{Compton profile}). Therefore, the differences between the effective Compton profiles and atomic Compton profiles quantify the deviations of our method from the former RIA treatments. Furthermore, the effective Compton profiles given in Eq. (\ref{effective Compton Profile}) and Eq. (\ref{effective Compton Profile1}) contain more dynamical information of the Compton scattering process, and depend on three variables: momentum component $p_{z}$, initial photon energy $\omega_{i}$, and scattering angle $\theta$. The atomic Compton profile, which is a single variable function of $p_{z}$, is totally determined by momentum distributions of atomic systems, irrespective of the dynamical properties of Compton scattering.

In this work, we use the differences between effective Compton profiles and atomic Compton profile to quantitatively describe the differences between our ``exact'' RIA results and the former RIA treatments. However, it must be noted that, the above-mentioned effective Compton profiles not only act as theoretical subjects but also can be directly measured from experiments. To experimentally determine these effective Compton profiles, we must first measure the differential cross section in real Compton scattering experiments, and then substitute them in Eq. (\ref{effective Compton Profile}) and Eq. (\ref{effective Compton Profile1}).

Before a detailed analysis of the effective Compton profiles, a briefing of the variable $p_{z}$ is required. As discussed in Section \ref{sec:2a}, the momentum component $p_{z}$ can be calculated in its exact form from Eq. (\ref{projection momentum}) or in its approximate from from Eq. (\ref{projection momentum2}). The momentum component $p_{z}$, when combined with two effective Compton profiles $\overline{J}_{\text{eff}}(p_{z},\omega_{i},\theta)$ and $J_{\text{eff}}(p_{z},\omega_{i},\theta)$ defined in Eq. (\ref{effective Compton Profile}) and Eq. (\ref{effective Compton Profile1}), produce four different effective Compton profiles. These four effective Compton profiles, on comparison with the conventional atomic Compton profile defined in Eq. (\ref{Compton profile}), can quantitatively describe the discrepancies between our ``exact'' RIA calculations with the four former RIA treatments presented in Fig. \ref{DDCS_figure1} and Fig. \ref{DDCS_figure2}, respectively.

The numerical results of effective Compton profiles at photon energy $\omega_{i}=356$ KeV and scattering angle $\theta=120^{\text{o}}$ are very similar to those of effective Compton profiles at $\omega_{i}=662$ KeV and $\theta=120^{\text{o}}$. Therefore, we only present the results which correspond to $\omega_{i}=662$ KeV and $\theta=120^{\text{o}}$ in Fig. \ref{ECP0} for C, Cu, Ge, and Xe atoms. In addition, the atomic Compton profiles computed using Eq. (\ref{Compton profile}) based on the nonrelativistic Hartree-Fock theory and the relativistic Dirac-Fock theory are presented for comparison. Earlier, Biggs \emph{et al.} calculated the atomic Compton profile using the nonrelativistic Hartree-Fock theory and the relativistic Dirac-Fock theory for light elements $Z<36$ and heavy elements $Z>36$, respectively \cite{Biggs}. To compare the nonrelativistic and relativistic results, we recalculate the atomic Compton profiles for C, Cu, and Ge atoms using the relativistic Dirac-Fock theory. We find that, for small-$Z$ element C with weak relativistic effect, no significant difference exists between nonrelativistic and relativistic results. However, for the middle-$Z$ elements Cu and Ge, the relativistic effects become stronger and obvious differences exist between nonrelativistic and relativistic results for large values of $|p_{z}|$. Moreover, to quantitatively analyse the discrepancies between the effective Compton profiles and the atomic Compton profiles, we define the relative discrepancy as follows:
\begin{equation}
D \equiv \frac{J_{\text{eff}}-J}{J}
\end{equation}
where $J$ and $J_{\text{eff}}$ represent the atomic Compton profile and the effective Compton profile, respectively. To equally consider the relativistic effects, we mark only the relative discrepancies between our RIA effective Compton profiles and relativistic atomic Compton profiles calculated using the Dirac-Fock theory. However, the relative discrepancies between the RIA effective Compton profiles and the nonrelativistic atomic Compton profiles, which are given by Biggs \emph{et al.} \cite{Biggs}, have the same order of magnitude.

An important observation can be made from this figure, which has significant importance in interdisciplinary studies, when the momentum component $|p_{z}|$ is less than 10 \emph{a.u.}, all the effective Compton profiles converge to the atomic Compton profiles with relative discrepancies $\mid D\mid < 20\%$. A non-negligible 20\% change of the variable $D$ arises only outside the palm of the momentum component $|p_{z}|$ greater than 10 \emph{a.u.}. Therefore, previous studies on condensed matter physics and material science, which studied electron correlations, electron momentum distributions, and Fermi surfaces using Compton profiles and Compton scattering experiments \cite{Kubo,Pisani,Cooper0,Cooper,Aguiar,Wang,Gillet,Sahariya,Rathor}, are still valid with a sufficiently high accuracy, because they are mainly focused on the region $|p_{z}|\sim$ \emph{a.u.}. However, in the large $|p_{z}|$ regions, except for the effective Compton profiles $\overline{J}_{\text{eff}}(p_{z},\omega_{i},\theta)$ defined in Eq. (\ref{effective Compton Profile}) that employ exact $p_{z}$ values, other effective Compton profiles have large discrepancies with the atomic Compton profiles in large momentum $|p_{z}|$ region, especially in the negative axis of $p_{z}$. The result is consistent with the conclusions obtained from the DDCS in Section \ref{sec:3a}, where large $|p_{z}|$ values corresponded to the cases where the final photon energy $\omega_{f}$ is far away from the Compton peak region. In these cases, our results of DDCS are notably different from those of the former RIA treatments. Another interesting phenomenon revealed by Fig. \ref{ECP0} is that, unlike the atomic Compton profiles, effective Compton profiles are generally not axisymmetric around the $p_{z}=0$ axis.

Furthermore, in the present work, we calculate all the effective Compton profiles at different initial photon energies $\omega_{i}$ and scattering angles $\theta$, and the results are presented in Appendix \ref{appendix3}. It is observable that the effective Compton profile $J_{\text{eff}}(p_{z},\omega_{i},\theta)$ defined in Eq. (\ref{effective Compton Profile1}) is more sensitive to the scattering angle $\theta$ than the incoming photon energy $\omega_{i}$. Moreover, the effective Compton profiles $J_{\text{eff}}(p_{z},\omega_{i},\theta)$ obtained for a smaller scattering angle $\theta$ has less discrepancy with the usual atomic Compton profiles; see Appendix \ref{appendix3} for more information.

\subsection{Numerical Uncertainty Estimate \label{sec:3c}}

To provide an uncertainty estimate for different numerical schemes, we recalculate the atomic Compton profiles by employing the same Hartree-Fock method as done by Biggs \emph{et al.} in reference \cite{Biggs}. The comparative results obtained for Ge and Xe atoms are provided in Table \ref{table} for selected $p_{z}$ momenta. The relative difference between our results and those in reference \cite{Biggs} provide an uncertainty estimate for different numerical schemes, and can be parameterized by the deviation parameter
\begin{equation}
D_{0} \equiv \mid \frac{J_{0}-J}{J_{0}} \mid
\end{equation}
where $J$ and $J_{0}$ correspond to the nonrelativistic atomic Compton profiles obtained from our calculations and Biggs \emph{et al.} \cite{Biggs}, respectively. This table clearly indicates that the uncertainties for different numerical schemes are in the range $10^{-4}$-$10^{-2}$, which is significantly less than the relative discrepancies $D$ between the atomic Compton profiles and the effective Compton profiles obtained in Section \ref{sec:3b}. Therefore, the uncertainties for different numerical schemes are neglected in this study.

\begin{table}
\caption{Comparative results of atomic Compton Profiles for Ge and Xe atoms. Our results and those of Biggs \emph{et al.} \cite{Biggs} are listed in the table. The deviation parameter is defined as $D_{0} \equiv \mid (J_{0}-J)/J_{0} \mid$ to characterize the uncertainties for different numerical schemes.}
\label{table}
\vspace{2mm}
\begin{ruledtabular}
\begin{tabular}{lccccccccc}
\multicolumn{4}{c}{Ge}
\\
\hline
$p_{z}$ (\emph{a.u.}) & Biggs \emph{et al.} & Our                   & deviation $D_{0}$
\\
\hline
0                     & 7.03                & 7.0439                & $2.0\times10^{-3}$
\\
0.5                   & 5.10                & 5.0960                & $7.9\times10^{-4}$
\\
1                     & 3.48                & 3.4721                & $2.3\times10^{-3}$
\\
2                     & 2.58                & 2.5832                & $1.2\times10^{-3}$
\\
5                     & $7.59\times10^{-1}$ & $7.5899\times10^{-1}$ & $8.3\times10^{-6}$
\\
10                    & $2.1\times10^{-1}$  & $2.1383\times10^{-1}$ & $1.8\times10^{-2}$
\\
20                    & $3.8\times10^{-2}$  & $3.8458\times10^{-2}$ & $1.2\times10^{-2}$
\\
40                    & $3.9\times10^{-3}$  & $3.8563\times10^{-3}$ & $1.1\times10^{-2}$
\\
\hline
\multicolumn{4}{c}{Xe}
\\
\hline
$p_{z}$ (\emph{a.u.}) & Biggs \emph{et al.} & Our                   & deviation $D_{0}$
\\
\hline
0                     & 9.74                & 9.7372                & $2.9\times10^{-4}$
\\
0.5                   & 8.21                & 8.2121                & $2.5\times10^{-4}$
\\
1                     & 5.45                & 5.4510                & $1.9\times10^{-4}$
\\
2                     & 3.68                & 3.6782                & $4.9\times10^{-4}$
\\
5                     & 1.30                & 1.2971                & $2.3\times10^{-3}$
\\
10                    & $5.1\times10^{-1}$  & $5.1505\times10^{-1}$ & $9.9\times10^{-3}$
\\
20                    & $1.2\times10^{-1}$  & $1.1723\times10^{-1}$ & $2.3\times10^{-2}$
\\
40                    & $1.8\times10^{-2}$  & $1.7538\times10^{-2}$ & $2.6\times10^{-2}$
\\
\end{tabular}
\end{ruledtabular}
\end{table}

\section{Conclusions and Perspectives \label{sec:4}}

In this study, we develop an ``exact'' numerical scheme to directly evaluate the integral to calculate the DDCS of the Compton scattering process in RIA formulation. Our method does not invoke any further simplified approximation or factorization treatment used in former RIA studies. The Compton scatterings for atomic systems are carefully analysed in this work, and our results are effectively compared with those of former treatments of RIA. Further, the validity of further simplified approximations and factorization results used in former RIA treatments can be tested using our approach. We select four typical elements C, Cu, Ge, and Xe in this study to represent the small-$Z$, middle-$Z$, and large-$Z$ regimes.

For the DDCS of Compton scatterings, our results agree well with those of former RIA treatments when $\omega_{f}\approx\omega_{C}$. Therefore, in the Compton peak region, where the RIA formulation is believed to be valid and reliable, the factorization treatments adopted in former RIA studies still hold, as Ribberfors \emph{et al.} expected earlier. However, when the scattered photon energy $\omega_{f}$ moves far away from the Compton peak region, notable discrepancies are observed. Some factorization treatments and kernel function approximations adopted in former RIA studies can produce large deviations in this region. In the entire energy spectrum, our results have little difference with the best of the former RIA treatments, which use the kernel function approximation $X(K_{i},K_{f}) \approx \overline{X}(p_{z})$ and employ the exact $p_{z}$ value. Furthermore, the comparisons with theoretical S-Matrix calculations and experimental results indicate that the available ranges of the RIA formulations are near the Compton peak. In regions far from the Compton peak, the RIA results become inaccurate, even when our ``exact'' numerical scheme is used.

To further quantitatively compare the differences between our ``exact'' RIA results and those of former RIA treatments, various effective Compton profiles are defined and calculated in this study. Detailed results indicate that, except for the effective Compton profile $\overline{J}_{\text{eff}}(p_{z},\omega_{i},\theta)$ defined in Eq. (\ref{effective Compton Profile}) by employing the exact $p_{z}$ values obtained from Eq. (\ref{projection momentum}), other effective Compton profiles exhibit notable discrepancies from atomic Compton profiles for large values of the momentum component $|p_{z}|$, especially in the negative axis of $p_{z}$. Furthermore, the following conclusions can be drawn from the analysis of effective Compton profiles for various incident photon energies $\omega_{i}$ and scattering angles $\theta$:

($i$). The effective Compton profiles do not show any notable difference from the atomic Compton profiles for a small momentum value $|p_{z}| < 10$ \emph{a.u.}. A non-negligible 20\% change of relative discrepancy $D$ arises only in the large momentum cases with $|p_{z}| > 10$ \emph{a.u.}. Therefore, the studies on condensed matter physics and material science that are focused on electron correlation, electron momentum distribution, and Fermi surfaces using Compton profiles and Compton scattering experiments, which correspond to $|p_{z}| \sim$ \emph{a.u.}, are still valid with sufficiently high accuracy.

($ii$). Unlike the atomic Compton profiles, the effective Compton profiles are generally not axisymmetric around the $p_{z}=0$ axis.

($iii$). The effective Compton profile $J_{\text{eff}}(p_{z},\omega_{i},\theta)$ defined in Eq. (\ref{effective Compton Profile1}) is more sensitive to the scattering angle $\theta$ than the incoming photon energy $\omega_{i}$. Moreover, the effective Compton profile $J_{\text{eff}}(p_{z},\omega_{i},\theta)$ obtained from a smaller scattering angle $\theta$ has less discrepancy with the usual atomic Compton profiles.

To summarize, in the present work, we conduct a comprehensive study of atomic Compton scatterings using our ``exact'' numerical treatment in the RIA formulation. Despite successfully employing the ``exact'' numerical evaluation for the integral in Eq. (\ref{doubly differential cross section1}) and not introducing any kernel function approximation and factorization treatment, our approach still relies on the physical picture of RIA, which is imperfect and has limitations. For example, the DDCS of Compton scattering in the RIA approach is started from Eq. (\ref{doubly differential cross section1}), which is, in addition, an approximation, and neglects a few interference terms in the dynamical process of Compton scattering. Some studies have indicated that the RIA approach can be realized by making leading order approximations for more advanced methods \cite{Eisenberger1,Pratt3,Pratt2}. In the past few years, several approaches beyond the IA formulation have been investigated \cite{Pratt,Pratt3,Pratt2,Drukarev0,Pratt5,Bergstrom,Pratt4,Kaplan,Suric,Drukarev1,Hopersky}. These works, which mainly employed low-energy theorems and S-matrix formulation, revealed many remarkable and nontrivial aspects of Compton scatterings and have gained a significant interests in interdisciplinary studies. Accordingly, we intend to study atomic Compton Scatterings beyond the IA formulation in the future.

\section*{ACKNOWLEDGMENTS}

We acknowledge helpful discussions with Qiang Du and Henry T. Wong. This work was supported by the National Natural Science Foundation of China (Grants No. 11475117, No. 11474209, No. 11975159 and No. 11975162),  the National Key Research and Development Program of China (Grant No. 2017YFA0402203) and the Fundamental Research Funds for the Central Universities.

\appendix

\section{Radial Hamiltonian and Bound State and Continuum State Orbitals \label{appendix0}}

In this Appendix, we give a detailed description on the Dirac orbitals we used in Section \ref{sec:2}. In this work, we only consider the spherical symmetric atomic systems. Therefore, the wavefunction of the electron state with definite quantum number $(n\kappa m)$, which is also called as the Dirac orbital, have the following form \cite{Desclaux,Grant,HuangSpin}
\begin{equation}
\psi_{n\kappa m}(\boldsymbol{r}) = \psi_{n\kappa m}(r,\theta,\phi)
                                 = \frac{1}{r}
                                   \left[
                                   \begin{array}{cc}
                                     G_{n\kappa}(r)\Omega_{\kappa m}(\theta,\phi) \\
                                     iF_{n\kappa}(r)\Omega_{-\kappa m}(\theta,\phi)
                                   \end{array}
                                   \right]
\end{equation}
where $G_{n\kappa}(r)$ and $F_{n\kappa}(r)$ are the large and small components respectively, $\Omega_{\kappa m}(\theta,\phi)$ is normalized spherical spinor defined as:
\begin{equation}
\Omega_{\kappa m}(\theta,\phi) = \sum_{s_{z}=\mu}\langle l m-\mu; \frac{1}{2} \mu| jm \rangle Y_{lm}(\theta,\phi) \chi_{\mu}
\end{equation}
where $Y_{lm}(\theta,\phi)$ is the spherical harmonics, $\langle l m-\mu; \frac{1}{2} \mu| jm \rangle$ is the Clebsch-Gordan coefficient, and $\chi_{\mu}$ is a spinor with $s=1/2$ and $s_{z}=\mu$.

In the above expressions, the quantum number $\kappa$ is the eigenvalue of operator
\begin{eqnarray}
K = \frac{1}{\hbar}
    \left(
      \begin{array}{cc}
        \boldsymbol{L}\cdot\boldsymbol{\alpha}+\hbar & 0 \\
        0 & -\boldsymbol{L}\cdot\boldsymbol{\alpha}-\hbar
      \end{array}
    \right)
\end{eqnarray}
with $\boldsymbol{L}$ to be orbital angular momentum vector and $\boldsymbol{\alpha}$ to be the conventional Dirac-$\alpha$ matrices. The information of orbital angular momentum $l$ and total momentum $j$ can be incorporated into quantum number $\kappa$ \cite{Sakurai}
\begin{eqnarray}
j & = & |\kappa|-\frac{1}{2}
\\
l & = & \bigg\{
        \begin{array}{cc}
          \kappa & \kappa > 0 \\
          -\kappa-1 & \kappa < 0
        \end{array}
\end{eqnarray}

In many cases, only the radial part need to be focused, and the angular part can be separated and neglected for simplicity. Therefore, we can introduce the following two-component radial Dirac orbitals:
\begin{equation}
u_{n\kappa}(r) \equiv u_{njl}(r) = \left[
                                   \begin{array}{c}
                                     G_{njl}(r) \\
                                     F_{njl}(r)
                                   \end{array}
                                   \right]
\label{bound state radial orbital}
\end{equation}

After introduction of the above Dirac orbitals, as well as its large and small components $F_{n\kappa}=F_{njl}$, $G_{n\kappa}=G_{njl}$, the Dirac-Fock equation for core and valence electrons can be expressed and solved routinely. Therefore, we can get the ground state wavefunctions for the whole atomic systems, as well as the wavefunctions for individual electrons.

\section{Equivalent Results for Numerical Integration \label{appendix1}}

In Section \ref{sec:2b}, we have mentioned that equivalent results for DDCS can be achieved by exchanging the order of integration in Eq. (\ref{doubly differential cross section2}). In this Appendix, we present results on differential cross sections calculated using alternative order of integration.

When we first integrate over momentum $p_{i}$ or polar angle $\theta_{1}$ in Eq. (\ref{doubly differential cross section2}), the DDCS for Compton scattering becomes
\begin{eqnarray}
\frac{d^{2}\sigma}{d\omega_{f}d\Omega_{f}} & = & \sum_{njl}
                                                 \frac{r_{0}^{2}}{2} \frac{\omega_{f}}{\omega_{i}} m^{2}c^{4}
                                                 \Theta(\omega_{i}-\omega_{f}-E_{njl}^{B})N_{njl} \nonumber
\\
                                           &   & \iint{\widetilde{p}_{i}^{2}\sin{\theta_{1}} d\theta_{1} d\phi_{1} }
                                                 \bigg(
                                                   (\phi_{njl}^{G}(\widetilde{p}_{i}))^{2}+(\phi_{njl}^{F}(\widetilde{p}_{i}))^{2}
                                                 \bigg) \nonumber
\\
                                           &   & \times \frac{X(K_{i}(\widetilde{p}_{i},\theta_{1}),K_{f}(\widetilde{p}_{i},\theta_{1}))}
                                                             {E_{i}(\widetilde{p}_{i})E_{f}(\widetilde{p}_{i})} \nonumber
\\
                                           &   & \times \bigg|
                                                   \frac{\widetilde{p}_{i}c^{2}}{E_{i}(\widetilde{p}_{i})}
                                                   -\frac{\widetilde{p_{i}}c^{2}+(A\sin{\theta_{1}}-B\cos{\theta_{1}})c}
                                                         {E_{f}(\widetilde{p}_{i})}
                                                 \bigg|^{-1} \nonumber
\\
\label{doubly differential cross section4*}
\end{eqnarray}
\begin{eqnarray}
\frac{d^{2}\sigma}{d\omega_{f}d\Omega_{f}} & = & \sum_{njl}
                                                 \frac{r_{0}^{2}}{2} \frac{\omega_{f}}{\omega_{i}} m^{2}c^{4}
                                                 \Theta(\omega_{i}-\omega_{f}-E_{njl}^{B})N_{njl} \nonumber
\\
                                           &   & \iint{p_{i} dp_{i} d\phi_{1} }
                                                 \bigg(
                                                   (\phi_{njl}^{G}(p_{i}))^{2}+(\phi_{njl}^{F}(p_{i}))^{2}
                                                 \bigg) \nonumber
\\
                                           &   & \times \bigg|
                                                 \frac{X(K_{i}(p_{i},\widetilde{\theta}_{1}),K_{f}(p_{i},\widetilde{\theta}_{1}))\times\sin{\widetilde{\theta}_{1}}}
                                                      {E_{i}(p_{i})\times(A\sin{\widetilde{\theta}_{1}}+B\cos{\widetilde{\theta}_{1}})c}
                                                 \bigg| \nonumber
\\
\label{doubly differential cross section4**}
\end{eqnarray}
where
\begin{eqnarray}
A & \equiv & \omega_{i}-\omega_{f}\cos{\theta}
\\
B & \equiv & \omega_{f}\sin{\theta}\cos(\phi-\phi_{1})
\end{eqnarray}

The fixed momentum $\widetilde{p}_{i}$ and polar angle $\widetilde{\theta}_{1}$ are calculated by solving the zeros of function $f$ in Eq. (\ref{function f}) respectively. After tedious calculations, $\widetilde{p}_{i}$ can be expressed as:
\begin{eqnarray}
\widetilde{p_{i}} & = & \frac{-MN\pm\sqrt{N^{2}+(M^{2}-1)m^{2}c^{4}}}{c(M^{2}-1)} \label{fixed momentum}
\end{eqnarray}
where
\begin{eqnarray}
M & \equiv & \frac{A\cos{\theta_{1}}-B\sin{\theta_{1}}}
                  {\omega_{i}-\omega_{f}}
\\
N & \equiv & \frac{\omega_{i}\omega_{f}(1-\cos{\theta})}
                  {\omega_{i}-\omega_{f}}
\end{eqnarray}
Similar to the case of $\widetilde{\phi}_{1}$ discussed in Section \ref{sec:2b}, only those which satisfy Eq. (\ref{fixed momentum}) and the inequality $\widetilde{p}_{i}\geq0$ simultaneously can be regarded as physical allowed values of $\widetilde{p}_{i}$.

The fixed polar angle $\widetilde{\theta}_{1}$ can be expressed through $\sin{\widetilde{\theta}_{1}}$ or $\cos{\widetilde{\theta}_{1}}$. The expression for $\sin{\widetilde{\theta}_{1}}$ and $\cos{\widetilde{\theta}_{1}}$ are:
\begin{eqnarray}
\sin{\widetilde{\theta}_{1}} & = & \frac{-BC\pm\sqrt{A^{2}(A^{2}+B^{2}-C^{2})}}{A^{2}+B^{2}} \label{fixed theta1}
\\
\cos{\widetilde{\theta}_{1}} & = & \frac{B}{A}\sin{\widetilde{\theta}_{1}}+\frac{C}{A} \label{fixed theta2}
\end{eqnarray}
where $C$ is defined as
\begin{eqnarray}
C & \equiv & \frac{E_{i}(p_{i})(\omega_{i}-\omega_{f})-\omega_{i}\omega_{f}(1-\cos{\theta})}
                  {p_{i}c}
\end{eqnarray}
Similarly, only those which satisfy the above expressions (\ref{fixed theta1})-(\ref{fixed theta2}) and the inequality $-1 \leq \sin{\widetilde{\theta}_{1}} \leq 1$ simultaneously are physically reasonable values of $\widetilde{\theta}_{1}$.

\section{Validity of the Approximation $X(K_{i},K_{f}) \approx X_{KN}$ and $X(K_{i},K_{f}) \approx \overline{X}(p_{z})$ \label{appendix}}

In this Appendix, we numerically study the kernel function $X(K_{i},K_{f})$ in the integrand of Eq. (\ref{doubly differential cross section1}). Particularly, we test the validity of the approximations
\begin{equation}
X(K_{i},K_{f}) \approx X_{KN}=\frac{\omega_{i}}{\omega_{f}}+\frac{\omega_{f}}{\omega_{i}}-\sin^{2}\theta
\end{equation}
and
\begin{eqnarray}
X(K_{i},K_{f}) & \approx & \overline{X}(p_{z}) \nonumber
\\
               &    =    & \frac{K_{i}(p_{z})}{K_{f}(p_{z})}
                           +\frac{K_{f}(p_{z})}{K_{i}(p_{z})} \nonumber
\\
               &         & +2m^{2}c^{2}
                            \bigg(
                              \frac{1}{K_{i}(p_{z})}-\frac{1}{K_{f}(p_{z})}
                            \bigg) \nonumber
\\
               &         & +m^{4}c^{4}
                            \bigg(
                              \frac{1}{K_{i}(p_{z})}-\frac{1}{K_{f}(p_{z})}
                            \bigg)^{2}
\end{eqnarray}
used in former treatments of RIA discussed in Section \ref{sec:2}. In order to quantitatively describe the differences between kernel function $X(K_{i},K_{f})$ and its Klein-Nishina value $X_{KN}$, or the differences between $X(K_{i},K_{f})$ and its ``averaged'' value $\overline{X}(p_{z})$, we can define the relative factors $\alpha$ and $\beta$ to be
\begin{eqnarray}
\alpha & \equiv & \frac{X(K_{i},K_{f})}{X_{KN}}
\\
\beta & \equiv & \frac{X(K_{i},K_{f})}{\overline{X}(p_{z})}
\end{eqnarray}
Here, the function $X(K_{i},K_{f})$ is calculated in Eq. (\ref{function X}), in general it depends on the initial and final photon energy $\omega_{i}$ and $\omega_{f}$, scattering angle $\theta$ ,as well as the electron pre-collision momentum $\boldsymbol{p}_{i}=(p_{i},\theta_{1},\phi_{1})$. While the ``averaged'' value $\overline{X}(p_{z})$ depends on momentum component $p_{z}$ obtained in Eq. (\ref{projection momentum}) or Eq. (\ref{projection momentum2}), irrespective of the transverse momentum in the $xy$ plane. The Klein-Nishina value $X_{KN}$ correspond to the special case where the electrons are at rest in the target, namely $\boldsymbol{p}_{i}=0$.

\begin{figure}
\centering
\includegraphics[width=0.485\textwidth]{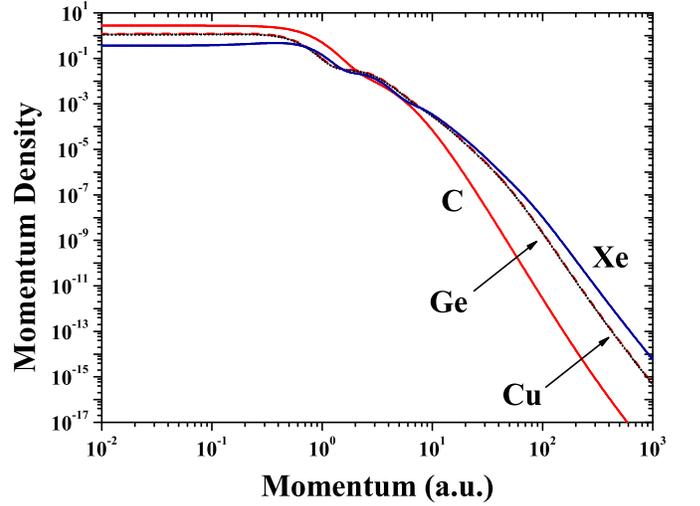}
\caption{Electron momentum distributions $\rho(p_{i})/Z$ for C, Ge, Cu and Xe atoms in the atomic units. It is worth noting that we have normalized the momentum distribution into the contribution from one electron.}
\label{Momentum_Distribution}
\end{figure}

Before a detailed analysis, we give the electron momentum distributions $\rho(p_{i})/Z$ for C, Cu, Ge and Xe atoms in Fig. \ref{Momentum_Distribution}. In this figure, momentum distribution of each element is normalized to give the contribution from one electron. The momentum distributions of Cu and Ge atoms are very similar to each other that can hardly be distinguished in the logarithmic coordinate. For all the elements C, Cu, Ge and Xe, momentum distributions decrease rapidly in large momentum region. Therefore, large momentum region gives negligible contributions on momentum distributions, compared with the small momentum region. Since the integrand in Eq. (\ref{doubly differential cross section1}) is proportional to the momentum distribution $\rho(p_{i})$, we can draw the conclusion that small momentum region plays a dominant role in calculating the DDCS of Compton scatterings from Eq. (\ref{doubly differential cross section1}), while large momentum region has tiny contributions in the integration. Contributions from large momentum region become notable only when the small momentum region is forbidden in the energy and momentum conservations.

\begin{figure}
\centering
\includegraphics[width=0.485\textwidth]{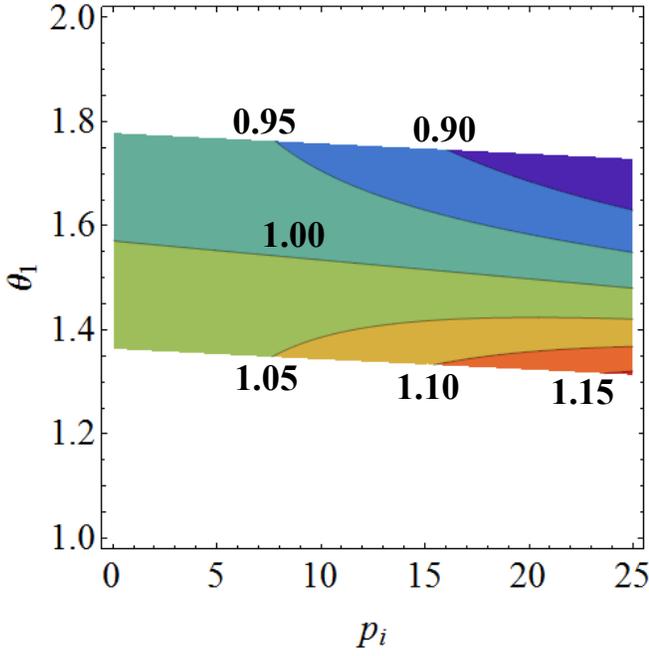}
\caption{Counterplot of relative factor $\alpha$ when outgoing photon energy $\omega_{f}$ goes in the Compton peak region. We select the following conditions: incoming photon energy $\omega_{i}=662$ KeV, outgoing photon energy $\omega_{f}=\omega_{C}=224.9$ KeV, and scattering angle $\theta=120^{\text{o}}$. The horizontal axis labels the electron momentum $p_{i}$ in units of \emph{a.u.}, and the vertical axis labels the polar angle of electron $\theta_{1}$ in units of radian. The regions which are not kinematically allowed in the energy and momentum conservations are left with white.}
\label{FunctionX1}
\end{figure}

\begin{figure}
\centering
\includegraphics[width=0.485\textwidth]{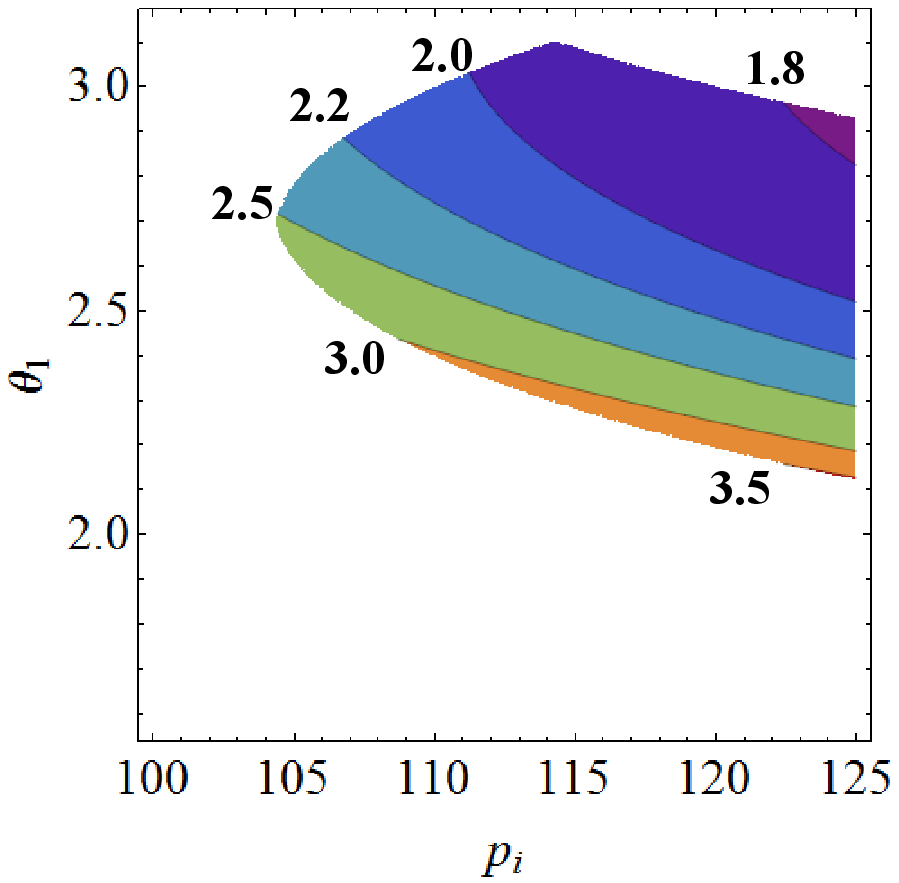}
\caption{Counterplot of relative factor $\alpha$ when final photon energy $\omega_{f}$ is far from the Compton peak region. We select the following conditions: initial photon energy $\omega_{i}=662$ KeV, final photon energy $\omega_{f}=500$ KeV, and scattering angle $\theta=120^{\text{o}}$. The horizontal and vertical axes label the electron momentum $p_{i}$ and polar angle $\theta_{1}$ similar to that in Fig. \ref{FunctionX1}. The regions which are not kinematically allowed in the energy and momentum conservations are left with white similar to that in Fig. \ref{FunctionX1}.}
\label{FunctionX2}
\end{figure}

\begin{figure}
\centering
\includegraphics[width=0.485\textwidth]{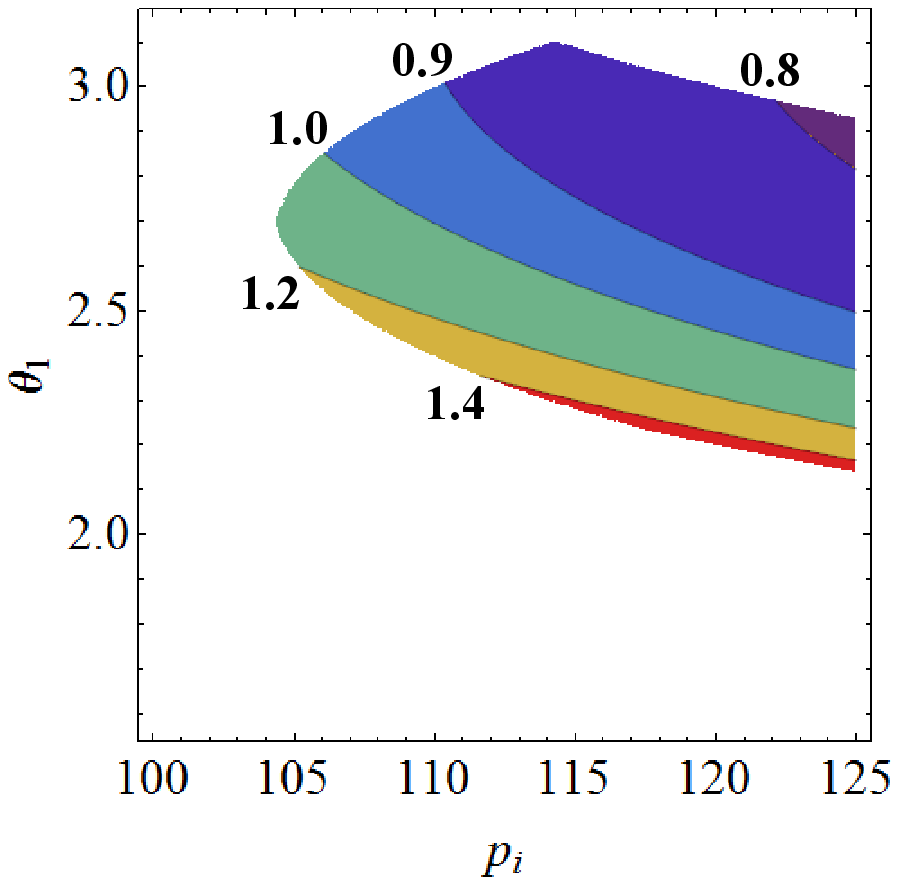}
\caption{Counterplot of relative factor $\beta$ when final photon energy $\omega_{f}$ is far from the Compton peak region. We select the following conditions: initial photon energy $\omega_{i}=662$ KeV, final photon energy $\omega_{f}=500$ KeV, and scattering angle $\theta=120^{\text{o}}$. The horizontal and vertical axes label the electron momentum $p_{i}$ and polar angle $\theta_{1}$ similar to that in Fig. \ref{FunctionX1}. The regions which are not kinematically allowed in the energy and momentum conservations are left with white similar to that in Fig. \ref{FunctionX1}.}
\label{FunctionX3}
\end{figure}

We select the case $\omega_{i}=662$ KeV and $\theta=120^{\text{o}}$ as a representative example, and present the numerical results of $\alpha$ and $\beta$ where final photon energy $\omega_{f}$ goes in and away from the Compton peak region, respectively. First, we consider the case correspond to the Compton peak region, where final photon energy is $\omega_{f}=\omega_{C}=224.9$ KeV. In this condition, the momentum component is $p_{z}=0$, which leads to $\overline{X}(p_{z})=X_{KN}$. Therefore, the relative factors $\alpha$ and $\beta$ coincide with each other, and only one of them need to be analysed quantitatively. The numerical values of factor $\alpha$ in this condition are shown in Fig. \ref{FunctionX1}. The minimal value of momentum that is kinematically allowed becomes $p_{i}^{\text{min}}=|p_{z}|=0$ \emph{a.u.}. In this figure we only plot the contributions from small momentum values $p_{i}=0-25$ \emph{a.u.}, where the momentum density is sufficiently large and can give notable contributions in the integration of Eq. (\ref{doubly differential cross section1}). This figure clearly shows that the relative factor $\alpha$ varies from $0.90-1.15$, which indicate the kernel function approximation $X(K_{i},K_{f}) \approx X_{KN}$ and $X(K_{i},K_{f}) \approx \overline{X}(p_{z})$ are valid and reliable in the Compton peak region. The result is consistent with conclusions obtained from Fig. \ref{DDCS_figure1} and Fig. \ref{DDCS_figure2}, which indicated that the results of our ``exact'' RIA treatment have small discrepancies with those of former RIA treatment in the Compton peak region $\omega_{f}\approx\omega_{C}$.

On the other hand, we select the conditions $\omega_{f}=500$ KeV and $\theta=120^{\text{o}}$ to illustrate the case where final photon energy is far away from the Compton peak region. The results of relative factors $\alpha$ and $\beta$ are given in Fig. \ref{FunctionX2} and Fig. \ref{FunctionX3}.  A crucial difference from the previous case is that small momentum values $p_{i}<100$ \emph{a.u.} are not kinematically allowed in energy and momentum conservations. Therefore, we only plot the region where electron momentum density has sufficiently large values. The minimal value of momentum that is kinematically allowed is $p_{i}^{\text{min}}=|p_{z}|\approx 105$ \emph{a.u.}. From these figures, we can observe that the relative factor $\alpha$ varies from $1.8-3.5$, while $\beta$ is between $0.8-1.4$, indicating that $X(K_{i},K_{f}) \approx \overline{X}(p_{z})$ is a better kernel function approximation than $X(K_{i},K_{f}) \approx X_{KN}$. This result is consistent with our DDCS results obtained from Fig. \ref{DDCS_figure1} and Fig. \ref{DDCS_figure2} in Section \ref{sec:3a}. The large momentum component $|p_{z}| \approx 105$ \emph{a.u.} corresponds to the cases where final photon energy is far away from the Compton peak region. In these conditions, the results of former RIA treatments agree with our ``exact'' RIA calculations only when $X(K_{i},K_{f}) \approx \overline{X}(p_{z})$ is used to calculate the DDCS of Compton scattering with exact $p_{z}$ values utilised. However, when $X(K_{i},K_{f}) \approx X_{KN}$ is adopted to calculate the DDCS, there are large discrepancies between our results and those of former RIA treatments, irrespective of whether exact or approximate $p_{z}$ values are employed.

\section{More Results of Effective Compton Profiles \label{appendix3}}

\begin{figure*}
\includegraphics[width=0.495\textwidth]{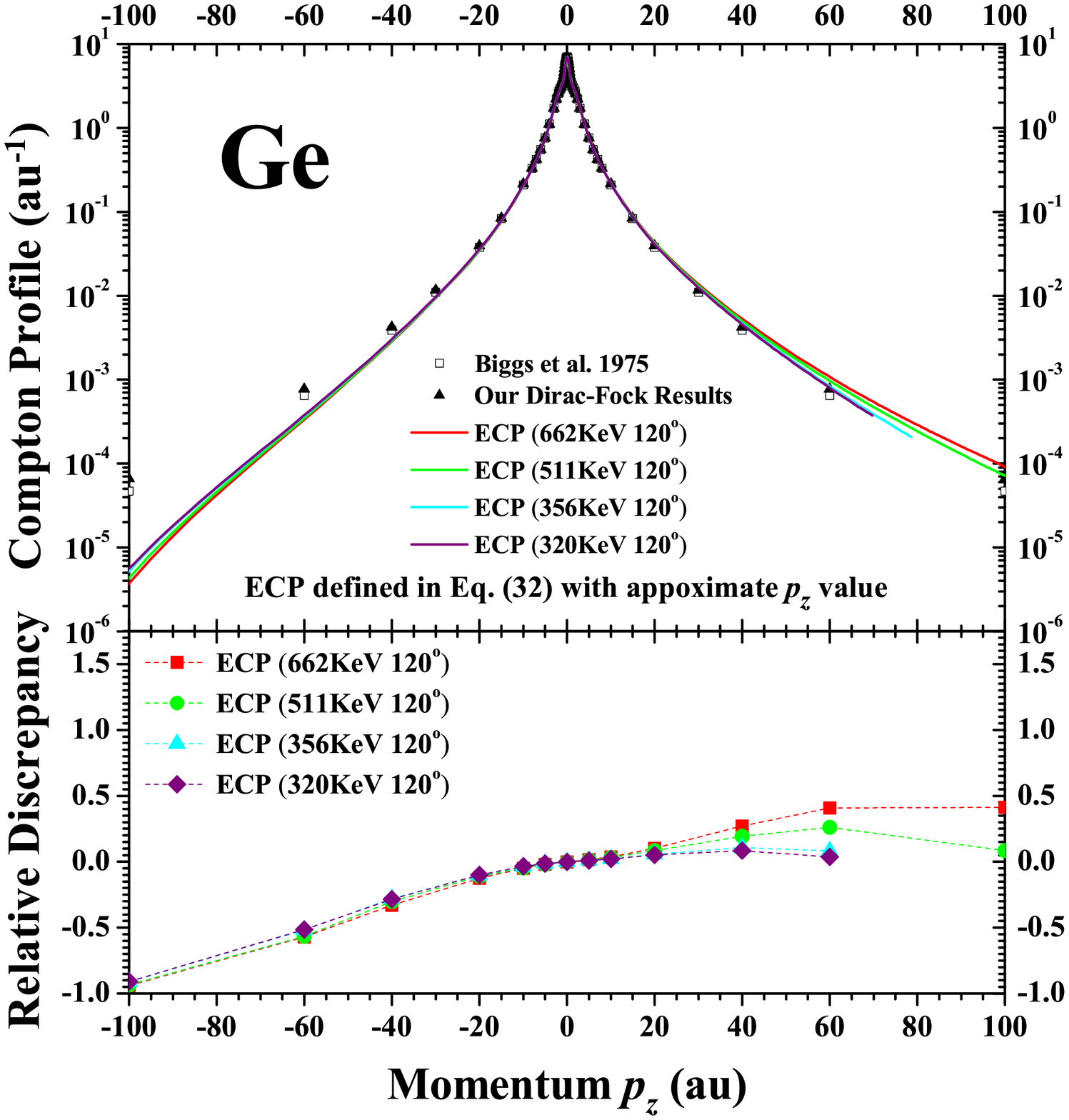}
\includegraphics[width=0.495\textwidth]{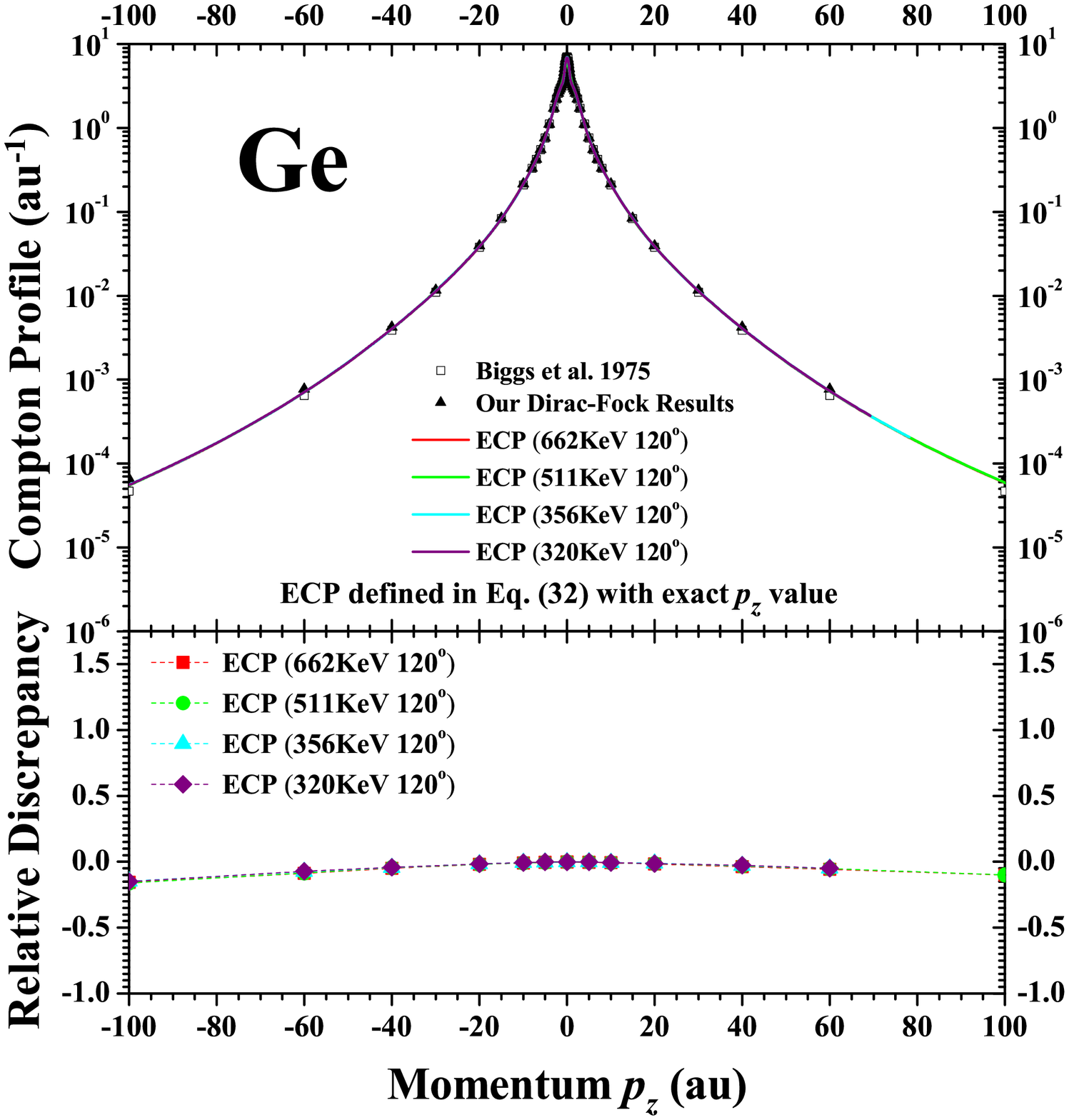}
\includegraphics[width=0.495\textwidth]{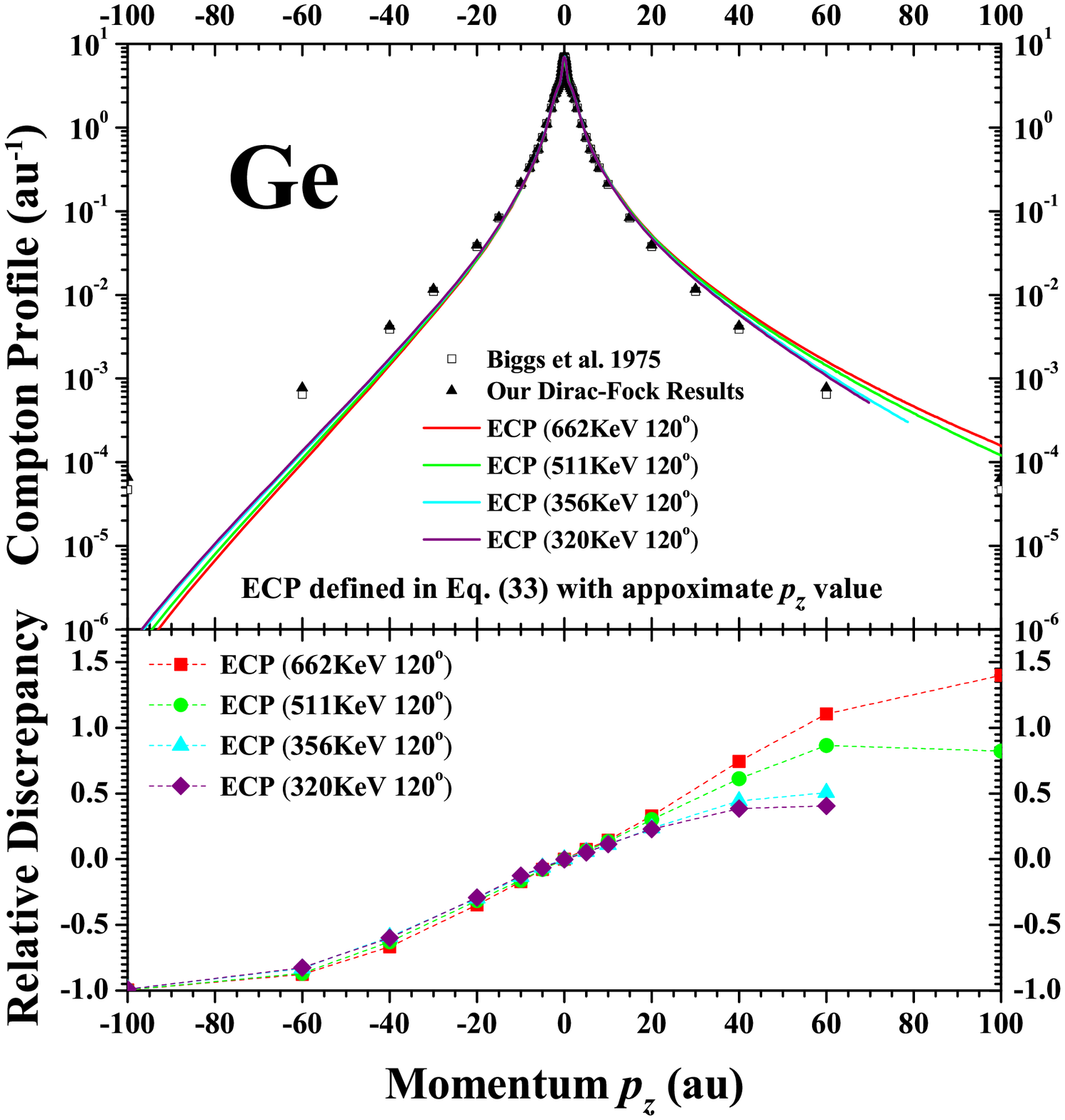}
\includegraphics[width=0.495\textwidth]{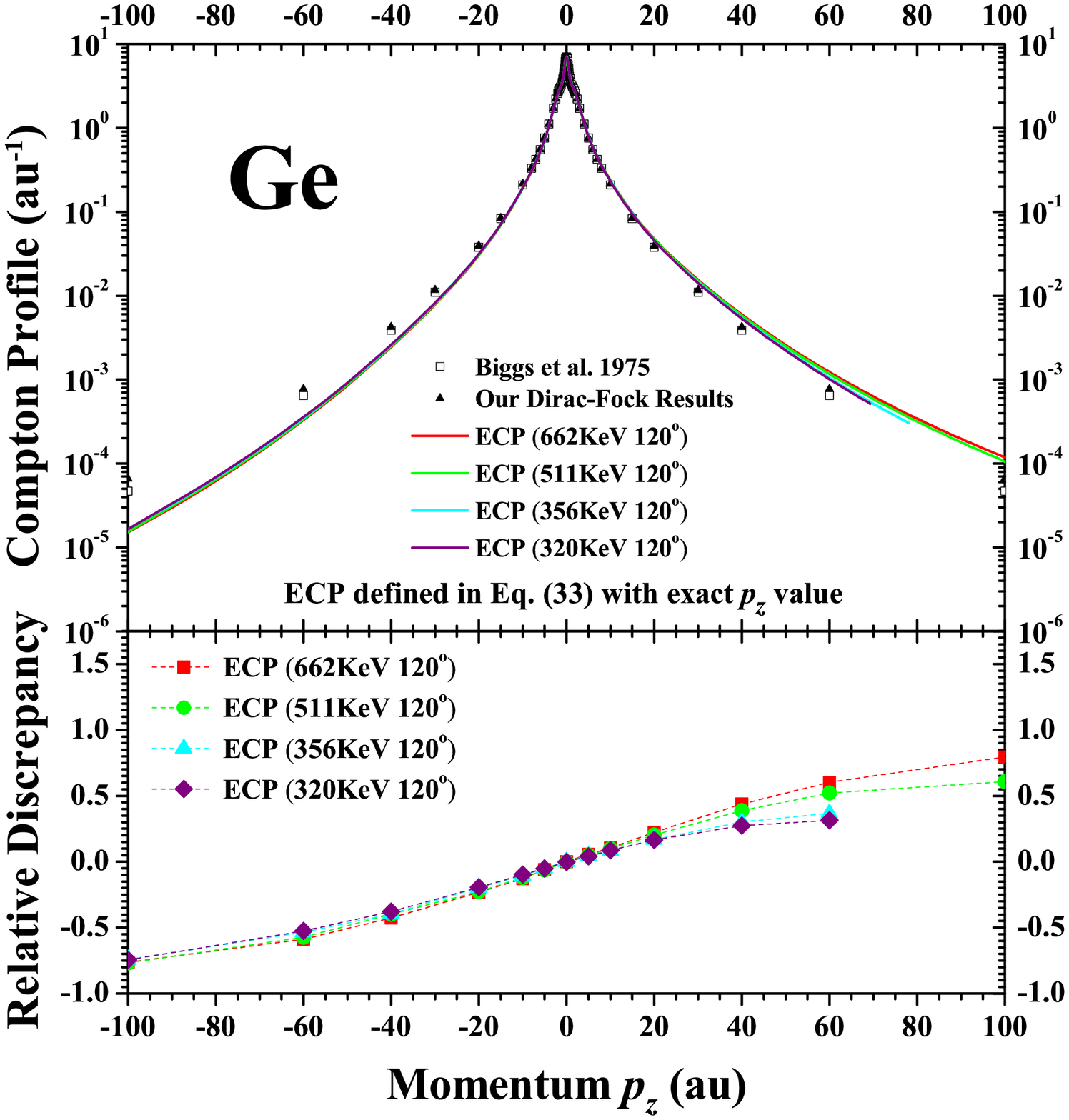}
\caption{Effective Compton profiles (ECP) of Ge atom at a scattering angle $\theta=120^{\text{o}}$ with various incident photon energies $\omega_{i}=662$ KeV, $511$ KeV, $356$ KeV, and $320$ KeV. Top left: effective Compton profile $\overline{J}_{\text{eff}}(p_{z},\omega_{i},\theta)$ defined in Eq. (\ref{effective Compton Profile}) with approximate $p_{z}$ values computed in Eq. (\ref{projection momentum2}). Top right: effective Compton profile $\overline{J}_{\text{eff}}(p_{z},\omega_{i},\theta)$ defined in Eq. (\ref{effective Compton Profile}) with exact $p_{z}$ values calculated in Eq. (\ref{projection momentum}). Bottom left: effective Compton profile $J_{\text{eff}}(p_{z},\omega_{i},\theta)$ defined in Eq. (\ref{effective Compton Profile1}) employing approximate $p_{z}$ values. Bottom right: effective Compton profile $J_{\text{eff}}(p_{z},\omega_{i},\theta)$ defined in Eq. (\ref{effective Compton Profile1}) employing exact $p_{z}$ values. Atomic Compton profiles $J(p_{z})$ computed using Eq. (\ref{Compton profile}) based on the nonrelativistic Hartree-Fock (HF) theory and the relativistic Dirac-Fock (DF) theory are plotted similar to Fig. \ref{ECP0}. Moreover, the relative discrepancies defined as $D\equiv (J_{\text{eff}}-J)/J$ are superimposed similar to that in Fig. \ref{ECP0}. It must be noted that for various effective Compton profiles, the momentum component $p_{z}$ has a maximal cut-off because of energy and momentum conservations. \label{ECP1}}
\end{figure*}

\begin{figure*}
\includegraphics[width=0.495\textwidth]{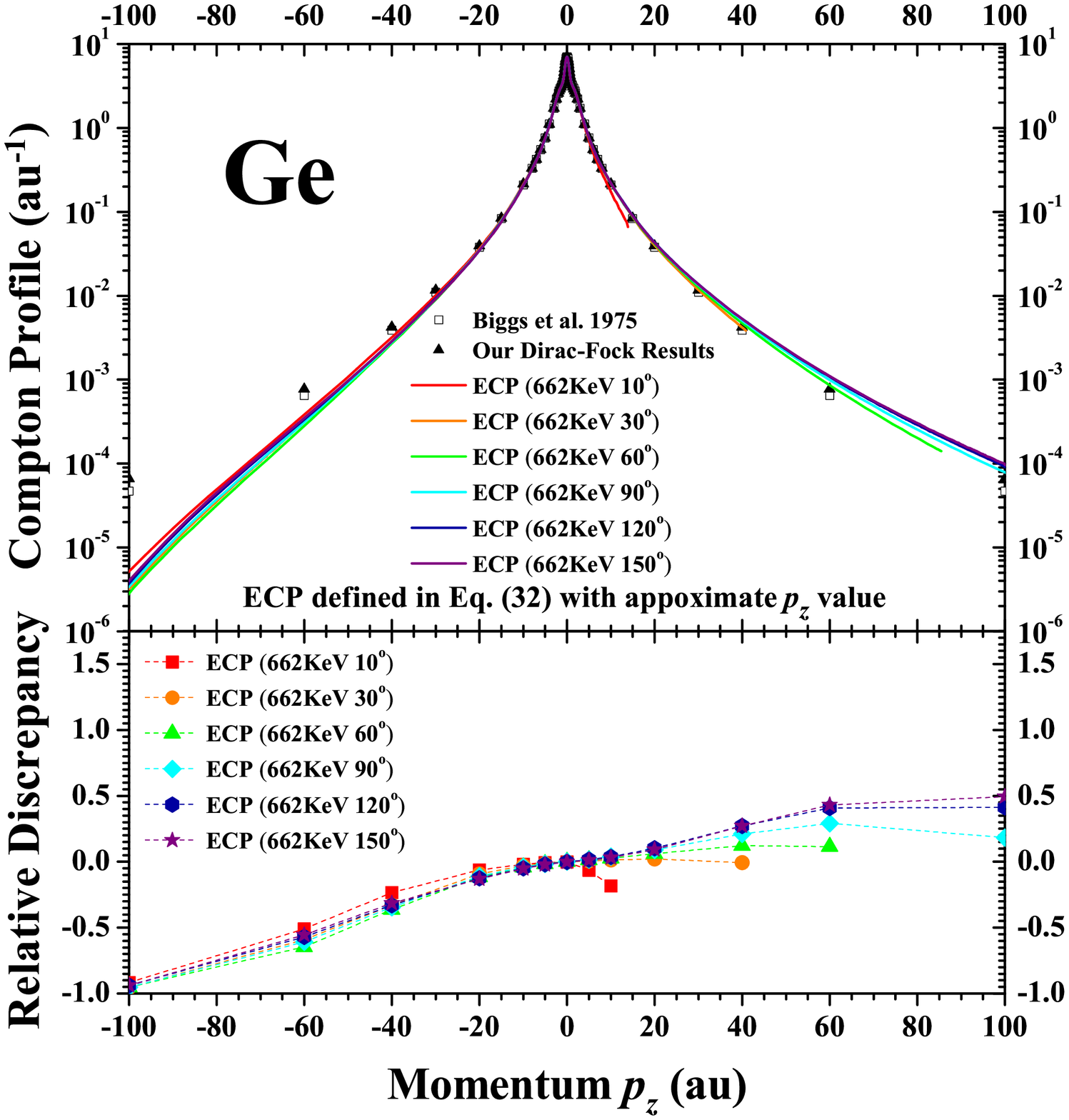}
\includegraphics[width=0.495\textwidth]{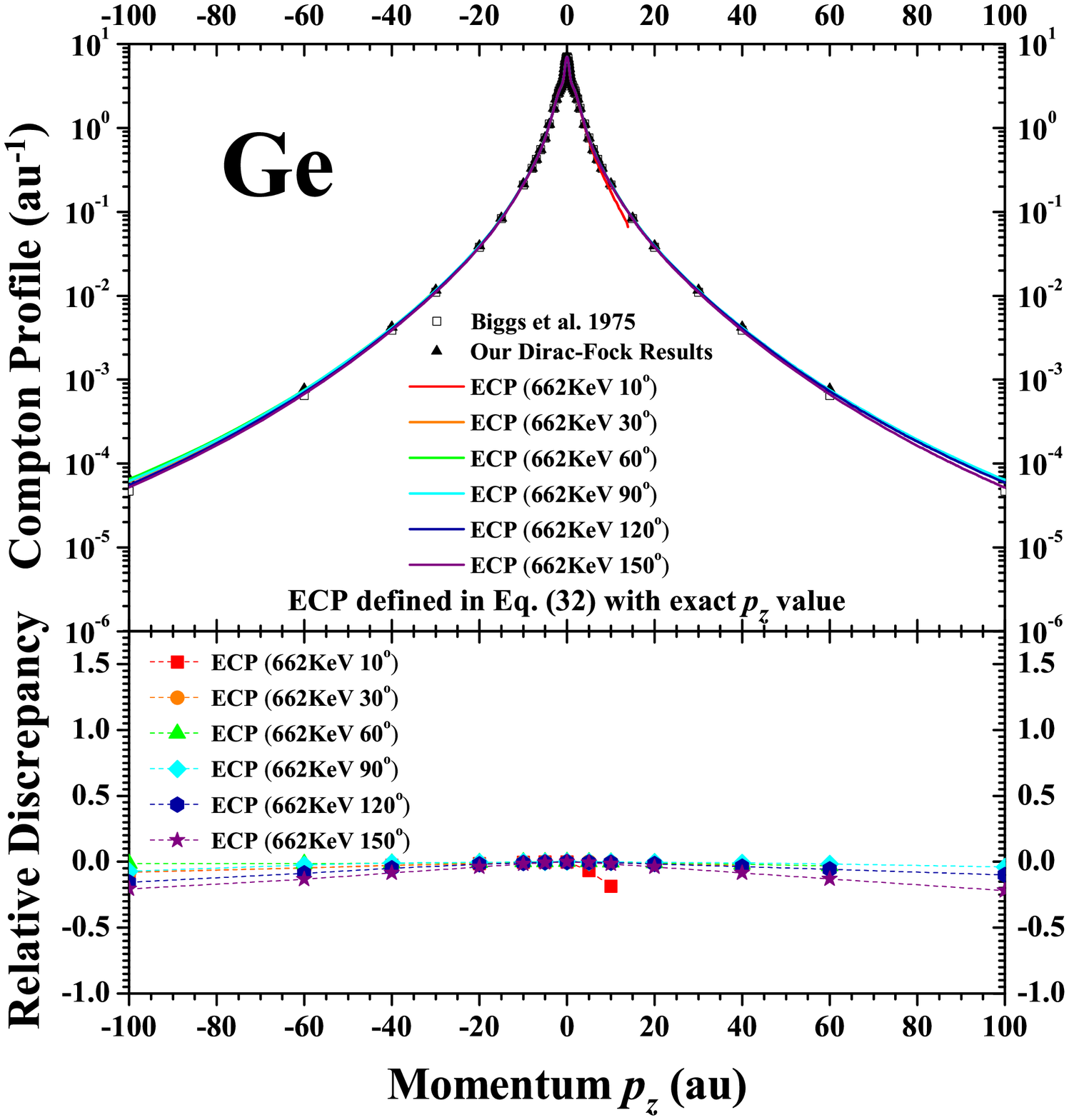}
\includegraphics[width=0.495\textwidth]{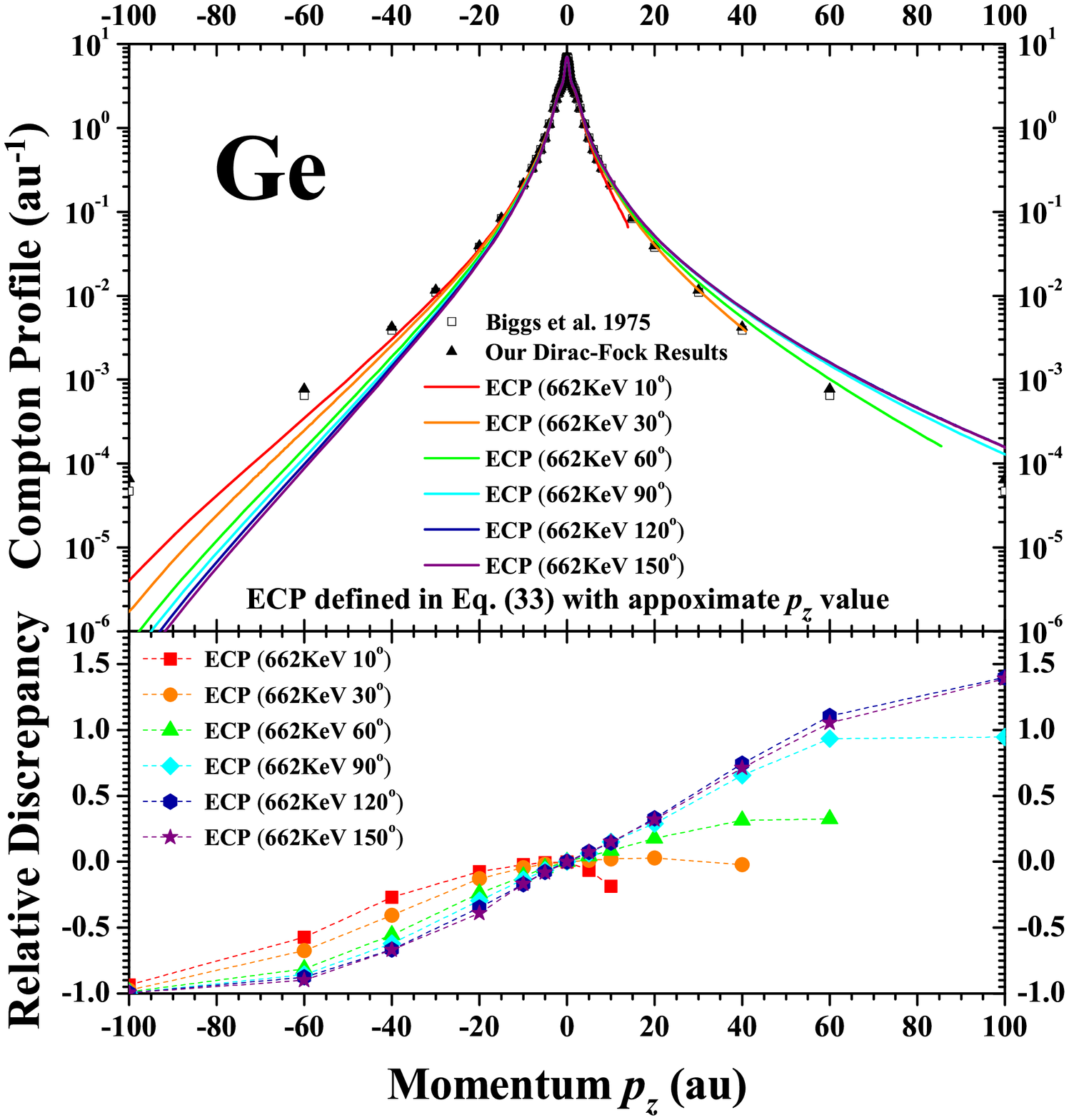}
\includegraphics[width=0.495\textwidth]{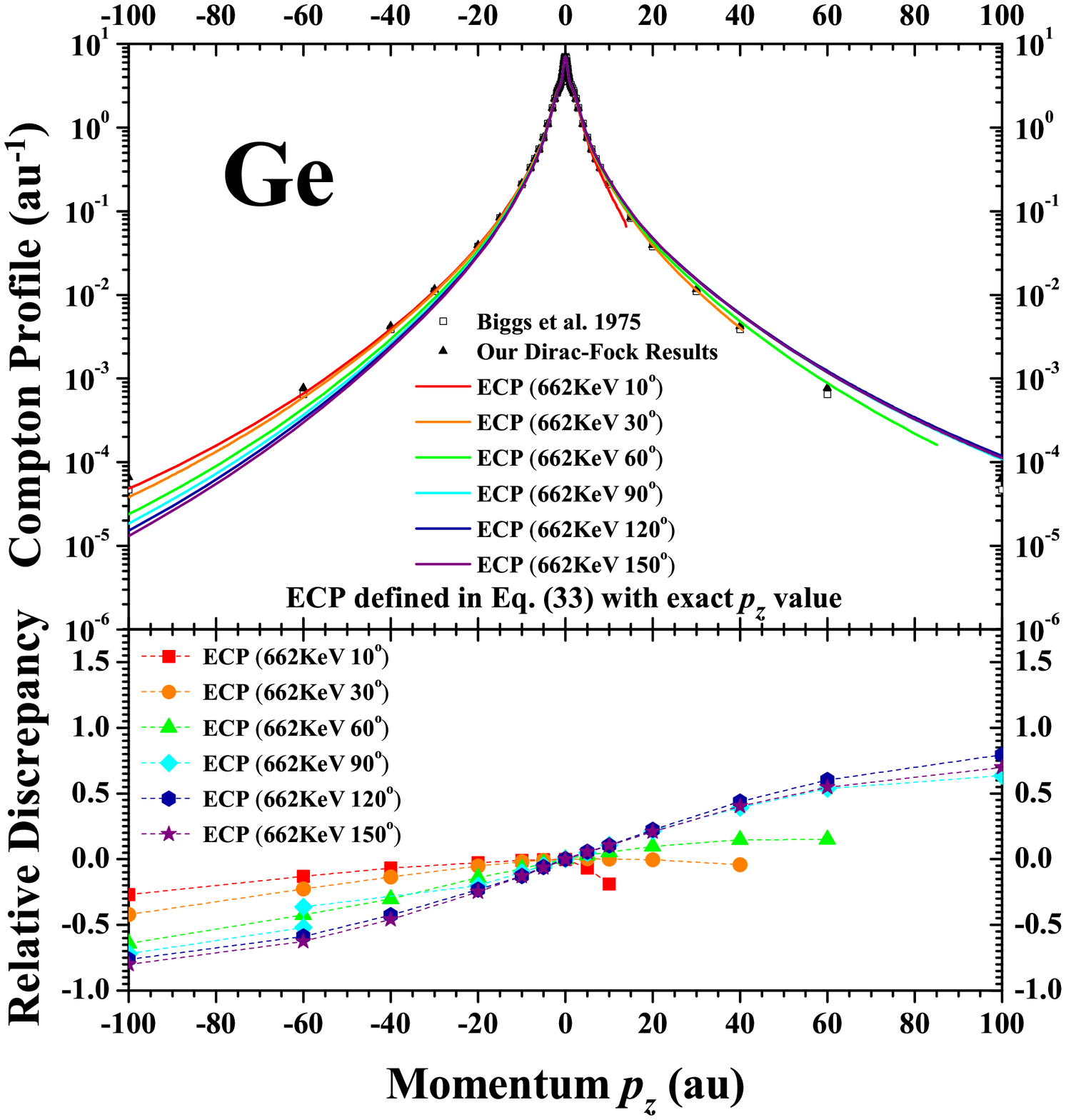}
\caption{Effective Compton profiles (ECP) of Ge atom at a photon energy $\omega_{i}=662$ KeV with various scattering angles $\theta=10^{\text{o}}$, $30^{\text{o}}$, $60^{\text{o}}$, $90^{\text{o}}$, $120^{\text{o}}$ and $150^{\text{o}}$. The effective Compton profiles $\overline{J}_{\text{eff}}(p_{z},\omega_{i},\theta)$ and $J_{\text{eff}}(p_{z},\omega_{i},\theta)$ defined in Eq. (\ref{effective Compton Profile}) and Eq. (\ref{effective Compton Profile1}) with approximate or exact $p_{z}$ values computed in Eq. (\ref{projection momentum2}) and Eq. (\ref{projection momentum}) are plotted similar to that in Fig. \ref{ECP1}. In addition, the atomic Compton profiles $J(p_{z})$ computed using Eq. (\ref{Compton profile}) based on nonrelativistic and relativistic theories are plotted in the figure. The relative discrepancies defined as $D\equiv (J_{\text{eff}}-J)/J$ are superimposed similar to that in Fig. \ref{ECP0} and Fig. \ref{ECP1}. It must be noted that for various effective Compton profiles, the momentum component $p_{z}$ has a maximal cut-off because of energy and momentum conservations. \label{ECP2}}
\end{figure*}

In this Appendix, we present more detailed numerical results of the effective Compton profiles $\overline{J}_{\text{eff}}(p_{z},\omega_{i},\theta)$ and $J_{\text{eff}}(p_{z},\omega_{i},\theta)$ defined in Section \ref{sec:3b}. These results reveal some intrinsic properties of effective Compton profiles. For simplicity, we list only the results of Ge atom. The results of C, Cu, and Xe atoms are similar to those of Ge.

The effective Compton profiles of Ge atom at a scattering angle $\theta=120^{\text{o}}$ with various incident photon energies are shown in Fig. \ref{ECP1}. In this work, we select several characteristic energies for gamma ray sources: $320$ KeV, $356$ KeV, $511$ KeV, and $662$ KeV. The effective Compton profiles $\overline{J}_{\text{eff}}(p_{z},\omega_{i},\theta)$ defined in Eq. (\ref{effective Compton Profile}) with approximate or exact $p_{z}$ values computed in Eq. (\ref{projection momentum2}) and Eq. (\ref{projection momentum}) are shown on the top. The effective Compton profiles $J_{\text{eff}}(p_{z},\omega_{i},\theta)$ defined in Eq. (\ref{effective Compton Profile1}) with approximate or exact $p_{z}$ values are shown at the bottom. In addition, the atomic Compton profiles computed using Eq. (\ref{Compton profile}) based on the nonrelativistic Hartree-Fock theory and the relativistic Dirac-Fock theory are presented for comparison. The relative discrepancies defined as $D\equiv (J_{\text{eff}}-J)/J$ in Section \ref{sec:3b} are superimposed similar to that in Fig. \ref{ECP0}. From this figure, we can observe that, only the effective Compton profile $\overline{J}_{\text{eff}}(p_{z},\omega_{i},\theta)$ defined in Eq. (\ref{effective Compton Profile}) which employs exact $p_{z}$ values fit well with the atomic Compton profiles for all values of $p_{z}$. Other effective Compton profiles are not axisymmetric around the $p_{z}=0$ axis and have large discrepancies at large $|p_{z}|$ values, specifically in the negative axis of $p_{z}$. It is worth noting that, for various effective Compton profiles, the momentum component $p_{z}$ has maximum and minimum values because of energy and momentum conservations in Compton scatterings. Moreover, for the same scattering angle $\theta$, when the incident photon energy $\omega_{i}$ is low, the maximal kinematically allowed value of $p_{z}$ becomes small. In all conditions, the minimum values of $p_{z}$ are less than $-100$ \emph{a.u.}, which are not shown in this figure. Similar to that in Section \ref{sec:3b}, when $|p_{z}| < 10$ \emph{a.u.}, all the effective Compton profiles are consistent with the atomic Compton profiles within $20\%$ uncertainty of the variable $D$. The deviations become pronounced only when $|p_{z}| > 10$ \emph{a.u.}, which corresponds to the cases where the final photon energy $\omega_{f}$ is far from the Compton peak region in the DDCS spectrum. Another interesting phenomenon is that all the effective Compton profiles at different energies $\omega_{i}$ almost converge with each other at a fixed scattering angle $\theta=120^{\text{o}}$.

The effective Compton profiles of Ge atom at photon energy $\omega_{i}=662$ KeV at various scattering angles $\theta$ are shown in Fig. \ref{ECP2}. We select the scattering angles $\theta=10^{\text{o}}$, $30^{\text{o}}$, $60^{\text{o}}$, $90^{\text{o}}$, $120^{\text{o}}$ and $150^{\text{o}}$ in this figure. The effective Compton profiles $\overline{J}_{\text{eff}}(p_{z},\omega_{i},\theta)$ and $J_{\text{eff}}(p_{z},\omega_{i},\theta)$ defined in Eq. (\ref{effective Compton Profile}) and Eq. (\ref{effective Compton Profile1}) with approximate or exact $p_{z}$ values computed in Eq. (\ref{projection momentum2}) and Eq. (\ref{projection momentum}) are plotted similar to Fig. \ref{ECP1}. Further, the nonrelativistic and the relativistic atomic Compton profiles are included in the figure. In addition, the relative discrepancies $D\equiv (J_{\text{eff}}-J)/J$ are superimposed similar to that in Fig. \ref{ECP0} and Fig. \ref{ECP1}. In these cases, the momentum $p_{z}$ has maximum and minimum values constrained by energy and momentum conservations, and the maximal kinematically allowed value of $p_{z}$ increases with the increase in scattering angle $\theta$. In these cases, only the effective Compton profile $\overline{J}_{\text{eff}}(p_{z},\omega_{i},\theta)$ defined in Eq. (\ref{effective Compton Profile}) calculated using exact $p_{z}$ values fit well with the atomic Compton profiles for all $p_{z}$ values. Other effective Compton profiles are not axisymmetric around the $p_{z}=0$ axis and show large discrepancies for large $|p_{z}|$ values. Fig. \ref{ECP2} clearly indicates that the effective Compton profiles $J_{\text{eff}}(p_{z},\omega_{i},\theta)$ defined in Eq. (\ref{effective Compton Profile1}) for different scattering angles $\theta$ do not converge with each other at a fixed incident photon energy $\omega_{i}=662$ KeV. However, Fig. \ref{ECP1} illustrates that the effective Compton profiles $J_{\text{eff}}(p_{z},\omega_{i},\theta)$ for different photon energies $\omega_{i}$ converge with each other at a fixed scattering angle $\theta=120^{\text{o}}$. Therefore, we can draw the conclusion that the effective Compton profile $J_{\text{eff}}(p_{z},\omega_{i},\theta)$ is more sensitive to scattering angle $\theta$ than incident photon energy $\omega_{i}$. Moreover, the effective Compton profile $J_{\text{eff}}(p_{z},\omega_{i},\theta)$ obtained from a smaller scattering angle has less discrepancy with the usual atomic Compton profiles. Furthermore, when $|p_{z}|$ is less than $10$ \emph{a.u.}, all the effective Compton profiles are consistent with the atomic Compton profiles within $20\%$ uncertainty of the variable $D$, similar to that in Fig. \ref{ECP0} and Fig. \ref{ECP1}. Therefore, previous studies on condensed matter physics relating to electron correlations, electron momentum distributions, and Fermi surfaces with Compton profiles are still valid with sufficiently high accuracy \cite{Kubo,Pisani,Cooper0,Cooper,Aguiar,Wang,Gillet,Sahariya,Rathor}.

\begin{figure*}
\centering
\includegraphics[width=0.75\textwidth]{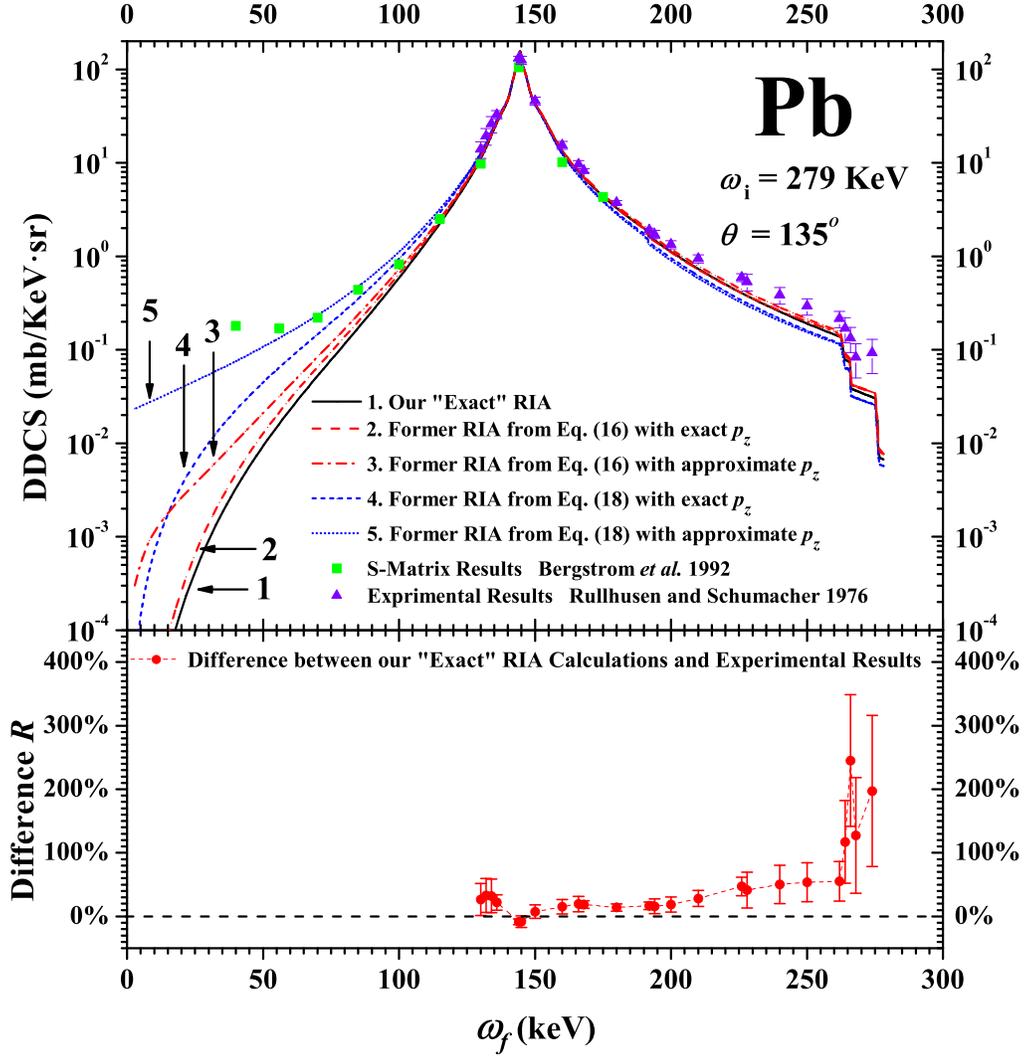}
\caption{Comparative results of DDCS of Compton scattering for Pb atom obtained from RIA and S-Matrix approaches. We select the following conditions: incident photon energy $\omega_{i}=279$ KeV and scattering angle $\theta=135^{\text{o}}$. Further, the experimental results of Rullhusen and Schumacher have been plotted in this figure. The differences between our ``exact'' RIA calculations and experimental results are presented at the bottom, and the experimental error bars of Rullhusen and Schumacher are also illustrated in this figure. The differences and the experimental error bars are expressed in percentage.}
\label{DDCS S Matrix}
\end{figure*}

\section{Comparison with S-Matrix Calculations \label{appendix5}}

In this Appendix, we present the comparative results of DDCS of Compton scattering obtained from the RIA formulation and the more advanced S-Matrix approach. These comparisons illustrate the available range for RIA formulation to address the atomic Compton scattering.

Fig. \ref{DDCS S Matrix} presents the comparative results for Pb atom at an incident photon energy $\omega_{i}=279$ KeV and a scattering angle $\theta=135^{\text{o}}$. The DDCS results for the Compton scattering process obtained using our ``exact'' RIA approach and several former RIA treatments are plotted similar to Fig. \ref{DDCS_figure1} and Fig. \ref{DDCS_figure2}. In addition, the S-Matrix and the experimental results are included in this figure. The S-Matrix calculations are performed by Bergstrom \emph{et al.} \cite{Bergstrom}, and the experimental results are measured by Rullhusen and Schumacher \cite{Rullhusen}. Furthermore, the differences between our ``exact'' RIA calculations and the experimental results, as well as experimental error bars, are presented in percentage at the bottom. The differences between theoretical calculations and experimental results are given by
\begin{equation}
R \equiv
\frac{ \bigg(\frac{d\sigma}{d\omega_{f}d\Omega_{f}}\bigg)_{\text{exp}} - \bigg(\frac{d\sigma}{d\omega_{f}d\Omega_{f}}\bigg)_{\text{theory}} }
     { \bigg(\frac{d\sigma}{d\omega_{f}d\Omega_{f}}\bigg)_{\text{theory}} }
\end{equation}

This figure indicates that the calculated results of our ``exact'' RIA treatment and former RIA treatments are in agreement with those of S-Matrix as well as with the experimental results in the Compton peak region. Considering the experimental uncertainties, only small differences exist between our ``exact'' RIA calculations and experimental results. However, both our ``exact'' RIA calculations and the former RIA treatments are inconsistent with the S-Matrix results outside the Compton peak region, specifically in the cases where the outgoing photon energy $\omega_{f}$ is very low. This could be because of the limitations of RIA formulations, which neglects some interference terms in the dynamical process of Compton scattering and is only a leading order approximation of the more advanced S-Matrix formalism \cite{Pratt,Pratt2}. The comparison presented here is consistent with the recent studies on Compton scattering, which have indicated that the available ranges for former RIA treatments are near the Compton peak \cite{Pratt,Kaliman,Pratt3,Pratt2,Drukarev0}. Because of the limited physical picture in RIA formulations, the reliable zones for our ``exact'' RIA treatments are still near the Compton peak. Our approach, despite employing exact numerical integration, does not exhibit a significant improvement over the former RIA treatments outside the Compton peak region.


\begin{thebibliography}{99}

\bibitem{Kubo}
    Y. Kubo, \emph{Electron correlation effects on Compton profiles of copper in the GW approximation}, J. Phys. Chem. Solids {\bf 66} 2202-2206 (2005).

\bibitem{Pisani}
    C. Pisani, M. Itou, Y. Sakurai, R. Yamaki, M. Ito, A. Erba, and L. Maschio, \emph{Evidence of instantaneous electron correlation from Compton profiles of crystalline silicon}, Phys. Chem. Chem. Phys. {\bf 13}, 933-936 (2011).

\bibitem{Cooper0}
    M. J. Cooper, \emph{Compton scattering and electron momentum distributions}, Adv. Phys. {\bf 20}, 453-491 (1971).

\bibitem{Cooper}
    M. J. Cooper, \emph{Compton scattering and the study of electron momentum density distributions}, Radiat. Phys. Chem. {\bf 50}, 63-76 (1997).

\bibitem{Aguiar}
    J. C. Aguiar, D. Mitnik, H. O. Di Rocco, \emph{Electron momentum density and Compton profile by a semi-empirical approach}, J. Phys. Chem. Solids {\bf 83}, 64-69 (2015).

\bibitem{Wang}
    Y. J. Wang, H. Lin, B. Barbiellini, P. E. Mijnarends, S. Kaprzyk, R. S. Markiewicz, and A. Bansil, \emph{Proposal to determine the Fermi-surface topology of a doped iron-based superconductor using bulk-sensitive Fourier-transform Compton scattering}, Phys. Rev. B {\bf 81}, 092501 (2010).

\bibitem{Pratt}
    P. M. Bergstrom Jr. and R. H. Pratt, \emph{An overview of the theories used in Compton Scattering Calculations}, Radiat. Phys. Chem. {\bf 50}, 3-29 (1997).

\bibitem{Porter}
    T. A. Porter, I. V. Moskalenko, A. W. Strong, E. Orlando, and L. Bouchet, \emph{Inverse Compton Origin of the Hard X-Ray and Soft Gamma-Ray Emission from the Galactic Ridge}, Astrophys. J {\bf 682}, 400-407 (2008).

\bibitem{Phuoc}
    K. Ta Phuoc, S. Corde, C. Thaury, V. Malka, A. Tafzi, J. P. Goddet, R. C. Shah, S. Sebban and A. Rousse, \emph{All-optical Compton gamma-ray source}, Nat. Photonics {\bf 6}, 308-311 (2012).

\bibitem{Takada}
    A. Takada \emph{et al.}, \emph{Development of an advanced Compton camera with gaseous TPC and scintillator}, Nucl. Insrum. Meth. A {\bf 546}, 258-262 (2005).

\bibitem{Mihailescu}
    L. Mihailescu, K. M. Vetter, M. T. Burks, E. L. Hull, and W. W. Craig, \emph{SPEIR: A Ge Compton camera}, Nucl. Insrum. Meth. A {\bf 570}, 89-100 (2007).

\bibitem{Chiu}
    J. -L. Chiu \emph{et al.}, \emph{The upcoming balloon campaign of the Compton Spectrometer and Imager (COSI)}, Nucl. Insrum. Meth. A {\bf 784}, 359-363 (2015).

\bibitem{Klein-Nishina}
    O. Klein and Y. Nishina, \emph{\"Uber die Streuung von Strahlung durch freie Elektronen nach der neuen relativistischen Quantendynamik von Dirac}, Z. Phys. {\bf 52},  853-868 (1929).

\bibitem{Sakurai}
    J. J. Sakurai, \emph{Advanced Quantum Mechanics} (Addison-Wesley, New-York, 1967).

\bibitem{Eisenberger1}
    P. Eisenberger and P. M. Platzman, \emph{Compton Scattering of X Rays from Bound Electrons}, Phys. Rev. A {\bf 2}, 415-423 (1970).

\bibitem{Eisenberger2}
    P. Eisenberger and W. A. Reed, \emph{Relationship of the relativistic Compton cross section to the electron's velocity distribution}, Phys. Rev. B {\bf 9}, 3237-3241 (1974).

\bibitem{Ribberfors1}
    R. Ribberfors, \emph{Relationship of the relativistic Compton cross section to the momentum distribution of bound electron states}, Phys. Rev. B {\bf 12}, 2067-2074 (1975).

\bibitem{Ribberfors2}
    R. Ribberfors, \emph{Relationship of the relativistic Compton cross section to the momentum distribution of bound electron states--II. Effects of anisotropy and polarization}, Phys. Rev. B {\bf 12}, 3136-3141 (1975).

\bibitem{Ribberfors3}
    R. Ribberfors, K.-F. Berggren, \emph{Incoherent-x-ray-scattering functions and cross sections $(d\sigma/d\Omega)_{incoh}$ by means of a pocket calculator}, Phys. Rev. A {\bf 26(6)}, 3325-3333 (1982).

\bibitem{Ribberfors4}
    R. Ribberfors,  \emph{X-ray incoherent scattering total cross sections and energy-absorption cross sections by means of simple calculation routines}, Phys. Rev. A {\bf 27}, 3061-3070 (1983); Erratum: Phys. Rev, A {\bf 28}, 2551 (1983).

\bibitem{Gillet}
    J.-M. Gillet, C. Fluteaux, and P. J. Becker, \emph{Analytical reconstruction of momentum density from directional Compton profiles}, Phys. Rev. B {\bf 60}, 2345-2349 (1999).

\bibitem{Sahariya}
    J. Sahariya and B. L. Ahuja, \emph{Compton profiles and electronic properties of Nd}, Phys. Scr. {\bf 84}, 065702 (2011).

\bibitem{Rathor}
    A. Rathor, V. Sharma, N. L. Heda, Y. Sharma, and B. L. Ahuja, \emph{Compton profiles and band structure calculations of IV-VI layered compounds GeS and GeSe}, Radiat. Phys. Chem. {\bf 77}, 391-400 (2008).

\bibitem{Brusa}
    D. Brusa, G. Stutz, J. A. Riveros, J. M. Fern\'andez-Varea, F. Salvat, \emph{Fast sampling algorithm for the simulation of photon Compton scattering}, Nucl. Insrum. Meth. A {\bf 379}, 167-175 (1996).

\bibitem{Salvat}
    Francesc Salvat and Jos\'e M Fern\'andez-Varea, \emph{Overview of physical interaction models for photon and electron transport used in Monte Carlo codes}, Metrologia {\bf 46}, S112-S138 (2009).

\bibitem{Brown}
    J. M. C. Brown, M. R. Dimmock, J. E. Gillam and D. M. Paganin, \emph{A low energy bound atomic electron Compton scattering model for Geant4}, Nucl. Instrum. Meth. B {\bf 338}, 77-88 (2014).

\bibitem{LaJohn}
    L. A. LaJohn, \emph{Low-momentum-transfer nonrelativistic limit of the relativistic impulse approximation expression for Compton-scattering doubly differential cross sections and characterization of their relativistic contributions}, Phys. Rev. A {\bf 81}, 043404 (2010).

\bibitem{Grant1961}
    I. P. Grant, \emph{Relativistic self-consistent fields} Proc. R. Soc. London Ser. A {\bf 262}, 555-576 (1961).

\bibitem{Desclaux1971}
    J. P. Desclaux, D. F. Mayersi and F. O \'Brien, \emph{Relativistic atomic wave functions} \emph{J. Phys. B: Atom. Molec. Phys.} {\bf 4} 631-642 (1971).

\bibitem{Desclaux}
    J. P. Desclaux, \emph{A multiconfiguration relativistic Dirac-Fock program}, Comput. Phys. Commun. {\bf 9}, 31-45 (1975).

\bibitem{Grant}
    K. G. Dyall, I. P. Grant, C. T. Johnson, F. A. Parpia and E. P. Plummer, \emph{GRASP: A general-purpose relativistic atomic structure program}, Comput. Phys. Commun. {\bf 55}, 425-456 (1989).

\bibitem{Ankudinov}
    A. L. Ankudinov, S. I. Zabinsky and J. J. Rehr, \emph{Single configuration Dirac-Fock atom code}, Comput. Phys. Commun. {\bf 98}, 359-364 (1996).

\bibitem{Visscher}
    L. Visscher and K. G. Dyall, \emph{Dirac-Fock atomic electronic structure calculations using different nuclear charge distributions}, At. Data and Nucl. Data Tables {\bf 67}, 207-224 (1996).

\bibitem{Undagoitia}
     T. M. Undagoitia and L. Rauch, \emph{Dark matter direct-detection experiments}, J. Phys. G: Nucl. Part. Phys. {\bf 43}, 013001 (2015).

\bibitem{CDEX0}
    S. K. Liu \emph{et al.} (CDEX Collaboration), \emph{Constraints on axion couplings from the CDEX-1 experiment at the China Jinping Underground Laboratory}, Phys. Rev. D {\bf 95}, 052006 (2017).

\bibitem{CDEX}
    H. Jiang \emph{et al.} (CDEX Collaboration), \emph{Limits on Light Weakly Interacting Massive Particles from the First 102.8kg $\times$ day Data of the CDEX-10 Experiment}, Phys. Rev. Lett. {\bf 120}, 241301 (2018).

\bibitem{CDMS}
    R. Agnese \emph{et al.} (SuperCDMS Collaboration), \emph{Results from the Super Cryogenic Dark Matter Search Experiment at Soudan}, Phys. Rev. Lett. {\bf 120}, 061802 (2018).

\bibitem{PandaX}
   X. Cui \emph{et al.} (PandaX-II Collaboration), \emph{Dark Matter Results from 54-Ton-Day Exposure of PandaX-II Experiment}, Phys. Rev. Lett. {\bf 119}, 181302 (2017).

\bibitem{LUX}
   D. S. Akerib \emph{et al.} (LUX Collaboration), \emph{Results from a Search for Dark Matter in the Complete LUX Exposure}, Phys. Rev. Lett. {\bf 118}, 021303 (2017).

\bibitem{XENON}
   E. Aprile \emph{et al.} (XENON Collaboration), \emph{Dark Matter Search Results from a One Ton-Year Exposure of XENON1T}, Phys. Rev. Lett. {\bf 121}, 111302 (2018).

\bibitem{Rodejohann}
    W. Rodejohann, \emph{Neutrinoless double-beta decay and neutrino physics}, J. Phys. G: Nucl. Part. Phys. {\bf 39}, 124008 (2012).

\bibitem{GERDA}
     M. Agostini \emph{et al.} (GERDA Collaboration), \emph{Results on Neutrinoless Double-$\beta$ Decay of $^{76}$Ge from Phase I of the GERDA Experiment}, Phys. Rev. Lett. {\bf 111}, 122503 (2013).

\bibitem{GERDA2}
     M. Agostini \emph{et al.} (GERDA Collaboration), \emph{Improved Limit on Neutrinoless Double-$\beta$ Decay of $^{76}$Ge from GERDA Phase II}, Phys. Rev. Lett. {\bf 120}, 132503 (2018).

\bibitem{EXO}
   J. B. Albert \emph{et al.} (EXO-200 Collaboration), \emph{Search for Majorana neutrinos with the first two years of EXO-200 data}, Nature {\bf 510}, 229-234 (2014).

\bibitem{KamLAND-Zen}
   A. Gando \emph{et al.} (KamLAND-Zen Collaboration), \emph{Search for Majorana Neutrinos Near the Inverted Mass Hierarchy Region with KamLAND-Zen}, Phys. Rev. Lett. {\bf 117}, 082503 (2016), Erratum: Phys. Rev. Lett. {\bf 117}, 109903 (2016).

\bibitem{Barker}
    D. Barker on behalf of the SuperCDMS Collaboration, \emph{Low Energy Background Spectrum in CDMSlite}, Proceedings of Science {\bf 282}, 874 (2017).

\bibitem{Ramanathan}
    K. Ramanathan, A. Kavner, A. E. Chavarria, P. Privitera, D. Amidei, T.-L. Chou, A. Matalon, R. Thomas, J. Estrada, J. Tiffenberg, and J. Molina, \emph{Measurement of low energy ionization signals from Compton scattering in a charge-coupled device dark matter detector}, Phys. Rev. D {\bf 96}, 042002 (2017).

\bibitem{Stutz}
    G. E. Stutz, \emph{Compton scattering cross section for inner-shell electrons in the relativistic impulse approximation} Nucl. Insrum. Meth. B {\bf 319}, 8-16 (2014).

\bibitem{Biggs}
    F. Biggs, L. B. Mendelsohn and J. B. Mann, \emph{Hartree-Fock Compton Profiles for the Elements}, At. Data and Nucl. Data Table {\bf 16}, 201-309 (1975).

\bibitem{Kaliman}
    Z. Kaliman, K. Pisk and R. H. Pratt, \emph{Compton scattering from positronium and validity of the impulse approximation}, Phys. Rev. A {\bf 83} 053406 (2011).

\bibitem{Pratt3} P. M. Bergstrom Jr., T. Suri\'c, K. Pisk, and R. H. Pratt, \emph{Compton scattering of photons from bound electrons: Full relativistic independent-particle-approximation calculations}, Phys. Rev. A {\bf 48}, 1134-1162 (1993).

\bibitem{Pratt2}
    R. H. Pratt, L. A. LaJohn, V. Florescu, T. Suri\'c, B. K. Chatterjee, S. C. Roy, \emph{Compton scattering revisited}, Radiat. Phys. Chem. {\bf 79}, 124-131 (2010).

\bibitem{Drukarev0}
    E. G. Drukarev and A. I. Mikhailov, \emph{High energy atomic physics}, Springer International Publishing: Switzerland (2016).

\bibitem{Pratt5}
    T. Suri\'c, P. M. Bergstrom Jr., K. Pisk, and R. H. Pratt, \emph{Compton scattering of photons by inner-shell electrons}, Phys. Rev. Lett. {\bf 67}, 189-192 (1991).

\bibitem{Bergstrom}
    P. M. Bergstrom, T. Suri\'c, K. Pisk and R. H. Pratt, \emph{Some preliminary calculations of whole atom Compton scattering of unpolarized photons} Nucl. Insrum. Meth. B {\bf 71}, 1-6 (1992).

\bibitem{Pratt4}
    M. Jung, R. W. Dunford, D. S. Gemmell, E. P. Kanter, B. Kr\"assig, T. W. LeBrun, S. H. Southworth, L. Young, J. P. J. Carney, L. LaJohn, R. H. Pratt, and P. M. Bergstrom Jr., \emph{Manifestations of Nonlocal Exchange, Correlation, and Dynamic Effects in X-Ray Scattering}, Phys. Rev. Lett. {\bf 81}, 1596-1599 (1998).

\bibitem{Kaplan} I. G. Kaplan, B. Barbiellini, and A. Bansil, \emph{Compton scattering beyond the impulse approximation}, Phys. Rev. B {\bf 68}, 235104 (2003).

\bibitem{Suric}
    T. Suri\'c, \emph{Compton scattering beyond impulse approximation: Correlation, nonlocal-exchange and dynamic effects}, Radiat. Phys. Chem. {\bf 75}, 1646-1650 (2006).

\bibitem{Drukarev1}
    E. G. Drukarev, A. I. Mikhailov and I. A. Mikhailov, \emph{Low-energy K-shell Compton scattering}, Phys. Rev. A {\bf 82}, 023404 (2010).

\bibitem{Hopersky}
    A. N. Hopersky, A. M. Nadolinsky, and S. A. Novikov, \emph{Compton scattering of two x-ray photons by an atom}, Phys. Rev. A {\bf 92}, 052709 (2015).

\bibitem{HuangSpin}   %% spin polarization formulism
   K.-N. Huang, \emph{Theory of angular distribution and spin polarization of photoelectrons}, Phys. Rev. A {\bf 22}, 223-239 (1980); Erratum: Phys. Rev. A {\bf 26}, 3676-3678 (1982).

\bibitem{Rullhusen}
   P. Rullhusen and M. Schumacher, \emph{Cross section profiles for Compton scattering of 279 2 keV photons by copper, tin and lead} J. Phys. B: Atom. Molec. Phys. {\bf 9}, 2435-2446 (1976).

\end{thebibliography}
\end{document}